\documentclass[twocolumn,longauth]{aa}
\usepackage{fixltx2e}
\usepackage[english]{babel}
\usepackage{graphicx,amsmath}
\usepackage{epsf,color} 
\usepackage[mathscr]{eucal}
\usepackage{amsmath}
\usepackage{amssymb,amsfonts}
\usepackage{natbib} 
\usepackage[varg]{txfonts}
\usepackage{dsfont}
\definecolor{Mygreen}{rgb}{0.00, 0.72, 0.0}
\definecolor{Mypink}{rgb}{1.0, 0.0, 0.5}
\usepackage[breaklinks, citecolor=blue, linkcolor=Mygreen, urlcolor=Mypink, colorlinks=true, debug, baseurl=' ']{hyperref}
\usepackage{float}
\usepackage{color}
\usepackage{widetext}
\usepackage{scalerel,stackengine,amsmath}

\bibpunct{(}{)}{;}{a}{}{,} 
\newcommand{\nc}{\newcommand}

\newcommand{\gobs}{G$_{0, \rm{obs}}$}
\newcommand{\CII}{[C\,{\sc ii}]}

\newcommand{\NII}{[N\,{\sc ii}]}

\newcommand{\OI}{[O\,{\sc i}]}

\newcommand{\HII}{H\,{\sc ii}}
\newcommand{\HI}{H\,{\sc i}}

\newcommand{\ergintensity}{\mbox{erg\,s$^{-1}$\,cm$^{-2}$\,sr$^{-1}$}}
\nc{\micron}{\mbox{$\mu$m}}
\nc{\cmcub}{\mbox{cm$^{-3}$}}
\nc{\cmsq}{\mbox{cm$^{-2}$}}
\nc{\Kkms}{\mbox{K~km~s$^{-1}$}}
\nc{\kms}{\mbox{km~s$^{-1}$}}
\nc{\mthirty}{\mbox{M\,33}}
\nc{\Tmb}{\mbox{$T_{\rm mb}$}}
\nc{\vlsr}{\mbox{v$_{\rm LSR}$}}
\nc{\twCO}{$^{12}$CO}
\nc{\thCO}{$^{13}$CO}
\nc{\msun}{\ensuremath{\mathrm{M}_\odot}}
\nc{\rsun}{\ensuremath{\mathrm{R}_\odot}}
\nc{\lsun}{\ensuremath{\mathrm{L}_\odot}}

\newcommand\equalhat{\mathrel{\stackon[1.5pt]{=}{\stretchto{%
    \scalerel*[\widthof{=}]{\wedge}{\rule{1ex}{3ex}}}{0.5ex}}}}

\begin{document}

\title{Gas and dust cooling along the major axis of M\,33 ({\tt
    HerM33es})\thanks{Herschel is an ESA space observatory with
    science instruments provided by European-led Principal
    Investigator consortia and with important participation from
    NASA.}\fnmsep\thanks{Maps of TIR, \CII, \OI\ shown in Figures 2, 3
    are available in electronic form at the CDS via anonymous ftp to
    cdsarc.u-strasbg.fr (130.79.128.5) or via
    http://cdsweb.u-strasbg.fr/cgi-bin/qcat?J/A+A/. }} \subtitle
      {Herschel/PACS \CII\ and \OI\ observations}

\titlerunning{Gas and dust cooling along the major axis of M\,33}


\author{
Carsten Kramer\inst{\ref{inst1},\ref{inst2}}
\and Thomas Nikola\inst{\ref{inst3}}
\and Sibylle Anderl\inst{\ref{inst5}, \ref{inst6}}
\and Frank Bertoldi\inst{\ref{inst7}}
\and M\'{e}d\'{e}ric Boquien\inst{\ref{inst8}}
\and Jonathan Braine\inst{\ref{inst9}}
\and Christof Buchbender\inst{\ref{inst10}}
\and Francoise Combes\inst{\ref{inst11}}
\and Christian Henkel\inst{\ref{inst12}, \ref{inst13}}
\and Israel Hermelo\inst{\ref{inst14}}
\and Frank Israel\inst{\ref{inst15}}
\and Monica Rela\~{n}o\inst{\ref{inst16}, \ref{inst22}}
\and Markus R\"ollig\inst{\ref{inst10}}
\and Karl Schuster\inst{\ref{inst1}}
\and Fatemeh Tabatabaei\inst{\ref{inst18}, \ref{inst21}}
\and Floris van der Tak\inst{\ref{inst19}, \ref{inst20}}
\and Simon Verley\inst{\ref{inst16}, \ref{inst22}}
\and Paul van der Werf\inst{\ref{inst15}}
\and Martina Wiedner\inst{\ref{inst11}}
\and Emmanuel M. Xilouris\inst{\ref{inst23}}
}

\institute{
%
  Institut de Radioastronomie Millim\'{e}trique (IRAM), 300 rue de la
  Piscine, 38406 Saint Martin d'H\`{e}res, France
  \email{kramer@iram.fr} \label{inst1} \and
  IRAM, Av. Divina Pastora 7, E-18012 Granada, Spain \label{inst2} \and
  Cornell University, Ithaca, NY 14852, USA \label{inst3} \and
  Frankfurter Allgemeine Zeitung, Hellerhofstra\ss{}e 2-4,
  60327 Frankfurt am Main, Germany \label{inst5} \and
  Univ. Grenoble Alpes, CNRS, IPAG, 38000 Grenoble, France \label{inst6} \and
  Argelander Institut f\"ur Astronomie. Auf dem H\"ugel 71, 
  D-53121 Bonn, Germany \label{inst7} \and
  Unidad de Astronom\'{i}a, Universidad de Antofagasta, Av. Angamos
  601, Antofagasta 1270300, Chile \label{inst8} \and
  Laboratoire d’Astrophysique de Bordeaux, Univ. Bordeaux, CNRS, B18N,
  allée Georoy Saint-Hilaire, 33615 Pessac, France \label{inst9} \and
  KOSMA, I. Physikalisches Institut, Universit\"at zu K\"oln,
  Z\"ulpicher Stra\ss{}e 77, D-50937 K\"oln, Germany \label{inst10} \and
  Observatoire de Paris, LERMA, College de France, CNRS, PSL Univ.,
  Sorbonne University, UPMC, Paris, France \label{inst11} \and
  Max Planck Institut f\"ur Radioastronomie, Auf dem H\"ugel 69,
  D-53121 Bonn, Germany \label{inst12} \and
  Department of Astronomy, King Abdulaziz University, PO Box 80203, 
  21589 Jeddah, Saudi Arabia \label{inst13} \and
  Instituto de Astrof\'{i}sica de Andaluc\'{i}a (IAA-CSIC), CAHA,
  Glorieta de la Astronom\'{i}a, s/n, 18008 Granada, Spain \label{inst14} \and
  Leiden Observatory, Leiden University, PO Box 9513,  
  NL 2300 RA Leiden, The Netherlands \label{inst15} \and
  Dept. F\'{i}sica Te\'{o}rica y del Cosmos, Universidad de
  Granada, 18012 Granada, Spain \label{inst16} \and
  Institute for Research in Fundamental Sciences-IPM, Larak Garden, 19395-5531 Tehran, Iran \label{inst18} \and 
  SRON Netherlands Institute for Space Research, Landleven 12, 9747
  AD Groningen, The Netherlands \label{inst19} \and
  Kapteyn Astronomical Institute, University of Groningen, The Netherlands \label{inst20} \and
  Instituto de Astrof\'{i}sica de Canarias, V\'{i}a L'actea S/N, E-38205 La
  Laguna, Spain \label{inst21} \and
  Instituto Universitario Carlos I de F\'{i}sica Te\'{o}rica y
  Computacional, Facultad de Ciencias, Universidad de Granada, E-18071
  Granada, Spain \label{inst22} \and
  Institute for Astronomy, Astrophysics, Space Applications \&
  Remote Sensing, National Observatory of Athens, P. Penteli,
  15236 Athens, Greece \label{inst23}
}

\date{Received / Accepted }

\abstract 
{M\,33 is a gas rich spiral galaxy of the Local Group. Its vicinity
  allows us to study its interstellar medium (ISM) on linear scales
  corresponding to the sizes of individual giant molecular clouds.}
{We investigate the relationship between the two major gas cooling lines
  and the total infrared (TIR) dust continuum. }
{We mapped the emission of gas and dust in M\,33 using the
  far-infrared lines of \CII\ and \OI(63\,$\mu$m) and the total
  infrared continuum.  The line maps were observed with the PACS
  spectrometer on board the Herschel Space Observatory.  These maps
  have 50\,pc resolution and form a $\sim370$\,pc wide stripe along
  its major axis covering the sites of bright \HII\ regions, but also
  more quiescent arm and inter-arm regions from the southern arm at
  $2\,$kpc galacto-centric distance to the south out to 5.7\,kpc
  distance to the north.  Full-galaxy maps of the continuum emission
  at $24\,\mu$m from Spitzer/MIPS, and at $70\,\mu$m, $100\,\mu$m, and
  $160\,\mu$m from Herschel/PACS were combined to obtain a map of the
  TIR. }
{ TIR and \CII\ intensities are correlated over more than two orders
  of magnitude. The range of TIR translates to a range of far
  ultraviolet (FUV) emission of \gobs$\sim 2$ to 200 in units of the
  average Galactic radiation field. The binned \CII/TIR ratio drops
  with rising TIR, with large, but decreasing scatter.  The
  contribution of the cold neutral medium to the \CII\ emission, as
  estimated from VLA \HI\ data, is on average only 10\%.  Fits of
  modified black bodies (MBBs) to the continuum emission were used to
  estimate dust mass surface densities and total gas column
  densities. A correction for possible foreground absorption by cold
  gas was applied to the \OI\ data before comparing it with models of
  photon dominated regions (PDRs). Most of the ratios of \CII/\OI\ and
  (\CII$+$\OI)/TIR are consistent with two model solutions. The median
  ratios are consistent with one solution at $n\sim2\cdot
    10^2$\,\cmcub, $G_0\sim60$, and and a second low-FUV solution at
    $n\sim10^4$\,\cmcub, $G_0\sim1.5$. }
%
{ The bulk of the gas along the lines-of-sight is represented by a
  low-density, high-FUV phase with low beam filling factors $\sim1$. A
  fraction of the gas may, however, be represented by the second
  solution.  }
%
%

\keywords{Galaxies: ISM -- Galaxies: individual: M33 -- Infrared:
  galaxies -- Infrared: ISM}
\maketitle

\section{Introduction}\label{sec:intro}

The strongest cooling line of the interstellar medium (ISM) in
galaxies is usually the \CII(158\,$\mu$m) line
\citep[e.g.,][]{brauher2008}. It is an important extinction-free
tracer of star formation. Only in regions of high density and high
temperature can the \OI\ 63\,$\mu$m line become the dominant coolant
\citep{kaufman1999}.  Mapping the emission of \CII\ and \OI\ in nearby
galaxies at high angular resolutions allows us to study their
variation with the local environment,
covering not only giant molecular clouds (GMCs) and sites of massive
star formation at different galacto-centric distances, but also the
diffuse inter-arm gas.

Galactic \CII\ emission originates from various ISM phases.  On global
scales in the Milky Way, 30\% of the \CII\ emission stems from
molecular gas which is bright in CO, that is from photon dominated
regions (PDRs), while 25\% arises from CO-dark molecular gas, 25\%
from atomic gas, and another 20\% from diffuse, ionized gas
\citep{pineda2014}. These contributions vary depending on the
environment and the location in the Milky Way. In a recent study of
ten active star-forming regions in the nearby galaxies NGC\,3184 and
NGC\,628, \citet{abdullah2017} showed that dense PDRs are the dominant
\CII\ emitters contributing $\sim70\%$, with other important
contributions from the warm ionized medium (WIM), while the atomic gas
only contributes with less than 5\% on average. They also conclude
that the relative strengths of all components vary significantly,
depending on the physical properties of the gas.

In general, \CII\ emission is well correlated with the star formation
rate (SFR) in galaxies. The correlation between SFR tracers derived
from H$\alpha$ and 24$\,\mu$m emission or from the total infrared
continuum (TIR), which is integrated between $1\mu$m and 1\,mm
wavelength, and \CII\ holds well on kiloparsec scales
\citep{herrera-camus2015}. However, on scales of $\sim50$\,pc, one
might expect to find a larger scatter as star-forming regions can be
spatially distinguished from the diffuse emission. The TIR traces the
emission of different grain size populations, from PAHs to large
grains, and in a variety of environments and phases. It measures the
dust heating caused by FUV photons, but also by soft optical photons
of insufficient energy to overcome the grain work function
\citep[E$<$6\,eV, ][]{tielens1985} and any Coulomb potential to eject
electrons, which heat the gas.  \citet{kapala2015} find that the
\CII/TIR ratio increases with galacto-centric distance in M\,31, and
\citet{kapala2017} show that this is caused by changes in the relative
hardness of the absorbed stellar radiation field, which are caused by
varying stellar populations, dust opacity, and galaxy metallicity,
while the photo-electric heating efficiency, the balance of the
photo-electric heating rate, and the grain FUV absorption rate
\citep{tielens2008} may stay constant.

The gas metallicity has a profound impact on the thermal balance of the
ISM and hence on the star formation. Low metallicity environments imply
lower dust abundance and a drop of the dust-to-gas ratio
\citep{draine2014}, allowing FUV photons of newborn OB stars to
penetrate more deeply into molecular clouds.
%

\object{M\,33}, also known as the Triangulum galaxy, is a gas-rich Sc
flocculent galaxy. With a total baryonic mass of
$\sim5\cdot 10^{10}$\msun\ \citep{corbelli2003} it is the third most
massive galaxy in the Local Group, after the Milky Way and
\object{M\,31} which are about a factor 20 more massive.  Its
proximity of 840\,kpc \citep{galleti2004} ($10''\equalhat$ 40.7\,pc)
makes it one of the nearest disk galaxies. It is only moderately
inclined by $56^\circ$ \citep{zaritsky1989}. Its metallicity is about
half solar \citep{ lin2017, toribio-san-cipriano2016, magrini2010},
similar to that of the Large Magellanic Cloud
\citep{hunter2007}. These properties make M\,33 an object of choice to
study the interplay of low-metallicity ISM and star formation on local
and global scales.


 M\,33 harbours several of the brightest giant \HII\ regions of the
 Local Group, including NGC\,604.  \citet{higdon2003} observed the six
 brightest \HII\ regions of M\,33 using ISO/LWS and its $70''$ beam
 (280\,pc). They concluded that more than half of the observed
 \CII\ arises from PDRs, and that the ionized gas lines can be modeled
 as arising from a single \HII\ component within their beams.  A cut
 along the major axis of M\,33 was studied by \citet[][]{kramer2013}
 using ISO/LWS data taken at galacto-centric distances from $-8$\,kpc
 to $+8$\,kpc, comparing these data with maps of CO and \HI\,
 integrated intensities.  They suggested that the fraction of
 \CII\ emission stemming from the cold neutral medium traced by \HI\
 rises from only 15\% in the inner $\pm4$\,kpc of M\,33 to 40\% in the
 outer parts. They also found the \CII/TIR ratio to rise from
 $\sim0.5$\% in the inner galaxy to about 4\% in the outer parts, at
 distances beyond 4\,kpc.

Here, we present maps of \CII, \OI\ (63$\mu$m), and the TIR taken
along the major axis of M\,33 using Herschel/PACS within the framework
of the {\tt HerM33es} key project \citep{kramer2010}.  Due to its
proximity, the spatial resolution provided by Herschel, $12''$ for the
\CII\ line corresponding to $\sim50$\,pc, allows us to probe the gas
and dust at the scale of individual GMCs \citep{gratier2017,
  tabatabaei2014, xilouris2012, boquien2011}. At these scales we
expect the global galactic conditions affecting GMCs to be constant
and the conditions of these clouds to be exclusively altered by the local
star formation.  It is then possible to study the star formation and
state of the ISM locally, but within the broader galactic
context. These maps include among others the southern arm, the nuclear
region, and several \HII\ regions, among them BCLMP\,691 and
\object{BCLMP\,302}. The PACS data of the latter region had been presented by
\citet{mookerjea2011} and they are included in the present work. The
two \HII\ regions BCLMP\,302 and BCLMP\,691 were also studied within
the {\tt HerM33es} project by \citet{mookerjea2016} and
\citet{braine2012}, respectively, using velocity-resolved spectra of
\CII\ obtained with HIFI in combination with spectra of \HI\, and CO,
discussing the presence of CO-dark molecular gas, and more generally,
trying to disentangle the contribution of the different ISM phases
along the lines-of-sight.

\begin{figure}[h!]
\centering
 \includegraphics[width=0.48\textwidth]{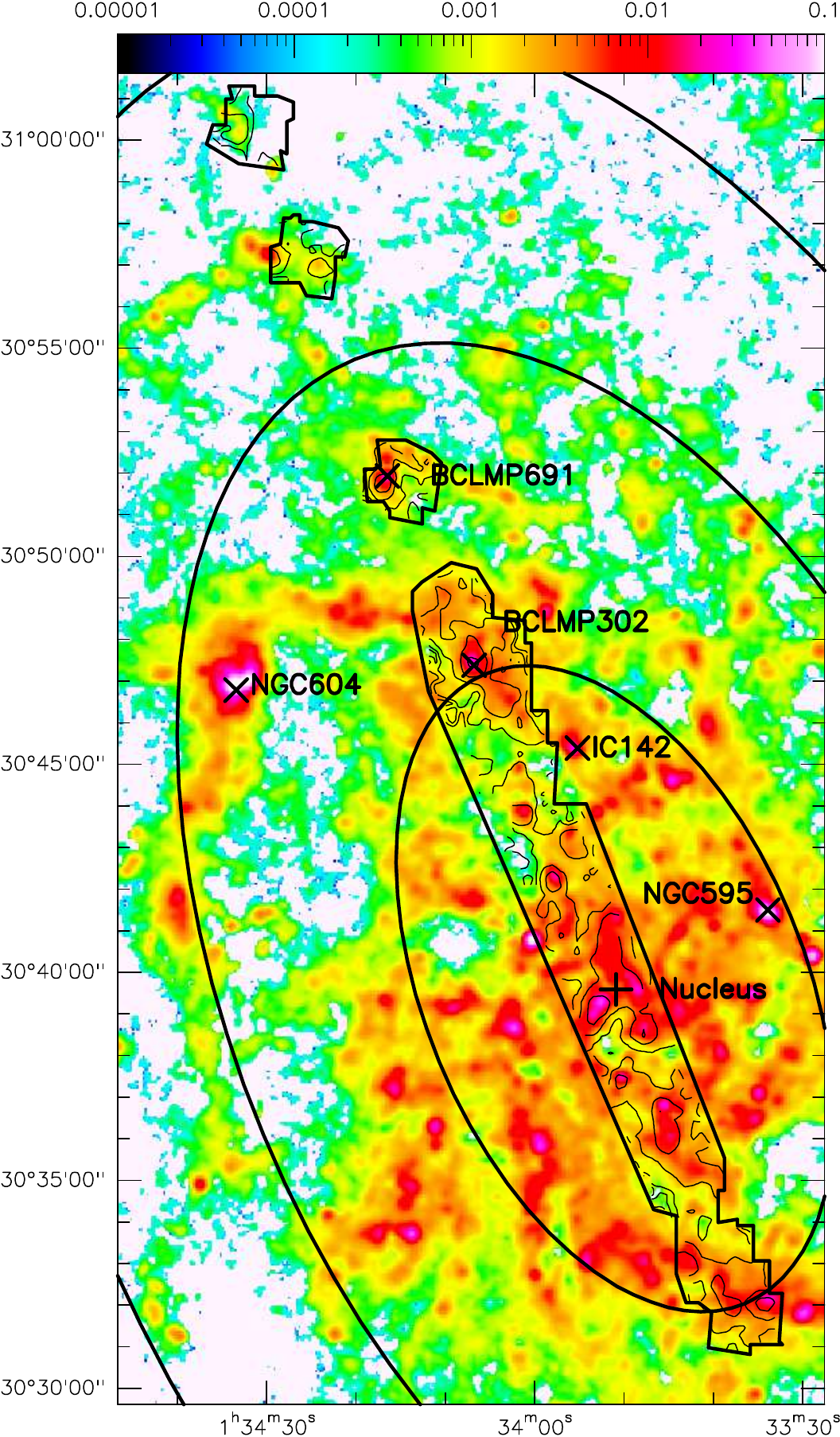}
 \caption{ M\,33 map of the total far-infrared (TIR) continuum in
   color together with \CII\ contours, both at $12''$ resolution.  TIR
   units are \ergintensity.  \CII\ contours are 3.4, 7.4, 16.0, 33.6,
   73.0 in units of $10^{-6}$\,\ergintensity. A polygon marks the
   outer edges of the regions mapped in \CII. Coordinates are in
   R.A. and Dec.  (eq. J2000). The nucleus at 1:33:50.9, 30:39:35.8
   (J2000) \citep{skrutskie.2006} is marked, together with a few
   prominent giant \HII\ regions. Ellipses delineate galacto-centric
   distances of 2, 4, and 6\,kpc. }
\label{fig:m33over}
\end{figure}

\section{Observations}\label{sec:obs}

\subsection{PACS spectroscopy}

\begin{figure*}
\centering
\includegraphics[scale=1.2]{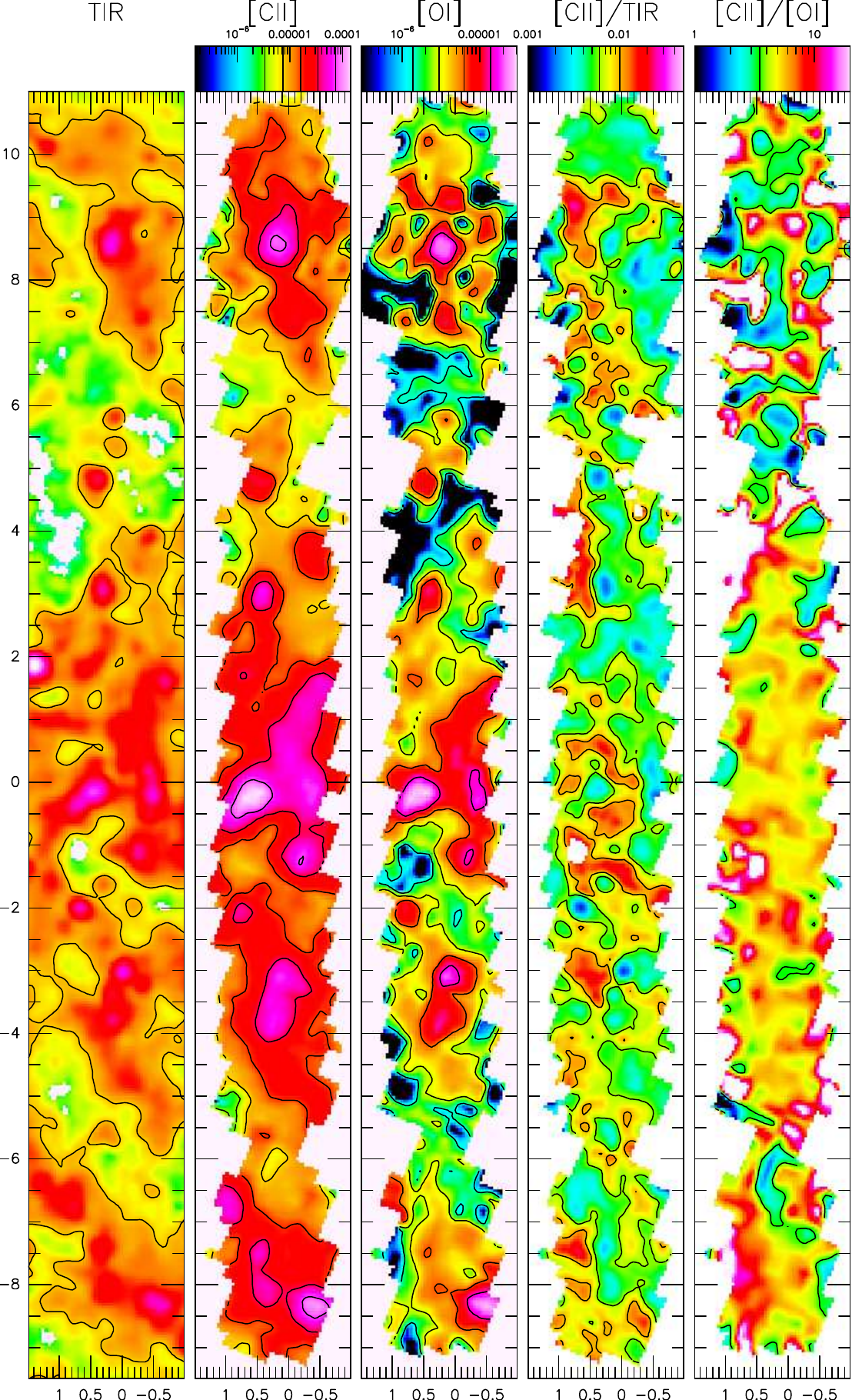}
 \caption{Central part of the PACS maps of M\,33. From left to right:
   intensities of total infrared continuum (TIR), \CII, \OI(63$\mu$m),
   \CII/TIR, and \CII/\OI\ intensity ratios.  Intensities are in units
   of \ergintensity.  Offsets are in arcminutes relative to the
   nucleus, after de-rotation of the position angle by
   22.5$^\circ$. The TIR contour is at $2\cdot 10^{-3}$\,\ergintensity,
   the threshold intensity used in the correlation plots to
   distinguish TIR-bright and TIR-weak regions
   (Figs.\,\ref{fig:tir-cii}, \ref{fig:ciitir-tir-r},
   \ref{fig:oivscii}, \ref{fig:ciioi}, \ref{fig:ciitir-tir-r}). The
   color wedge is the same as in
   Fig.\,\ref{fig:m33over}. \CII\ contours are the same as in
   Fig.\,\ref{fig:m33over}.  \OI\ contours are 1.4, 3.0, 6.3, 13.8 in
   units of $10^{-6}$\,\ergintensity. The map of \CII/TIR is overlayed
   with contours at the 0.6\% and 1\% level of the ratio.  The map of
   \CII/\OI\ is overlayed with the contour of 3.5. Long tic marks in
   the color wedges indicate contour levels. }
\label{fig:strips-cen}
\end{figure*}

\begin{figure*}
\centering
 \includegraphics[scale=0.9]{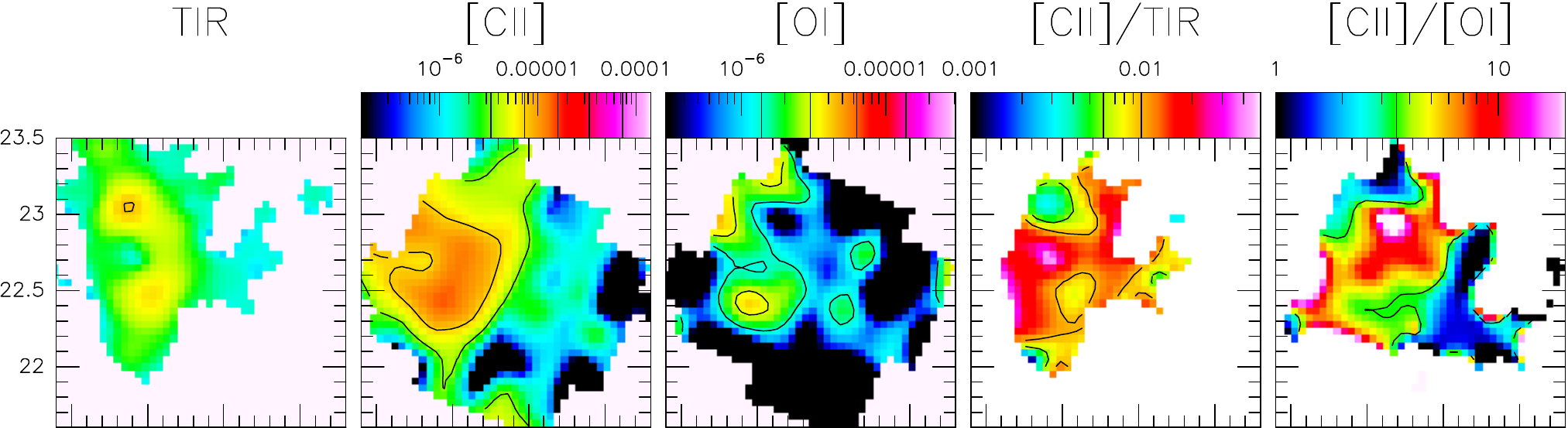}
 \includegraphics[scale=0.95]{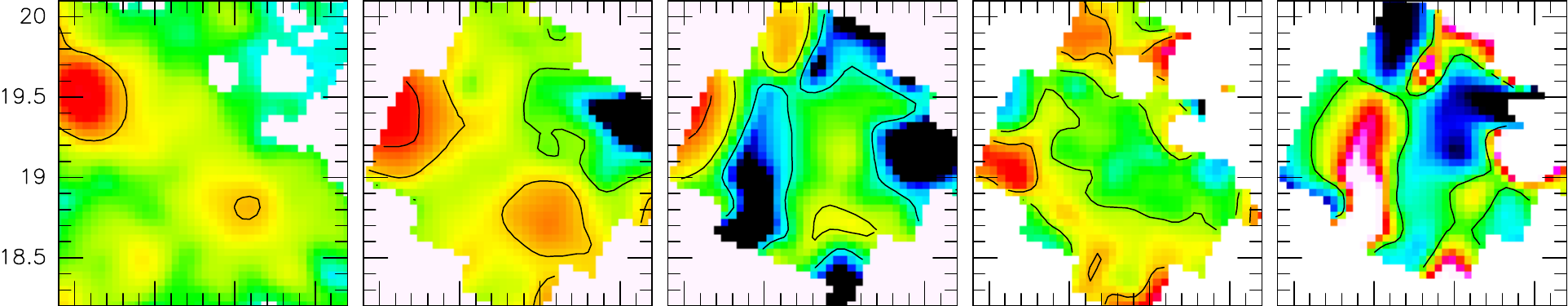}
 \includegraphics[scale=0.9]{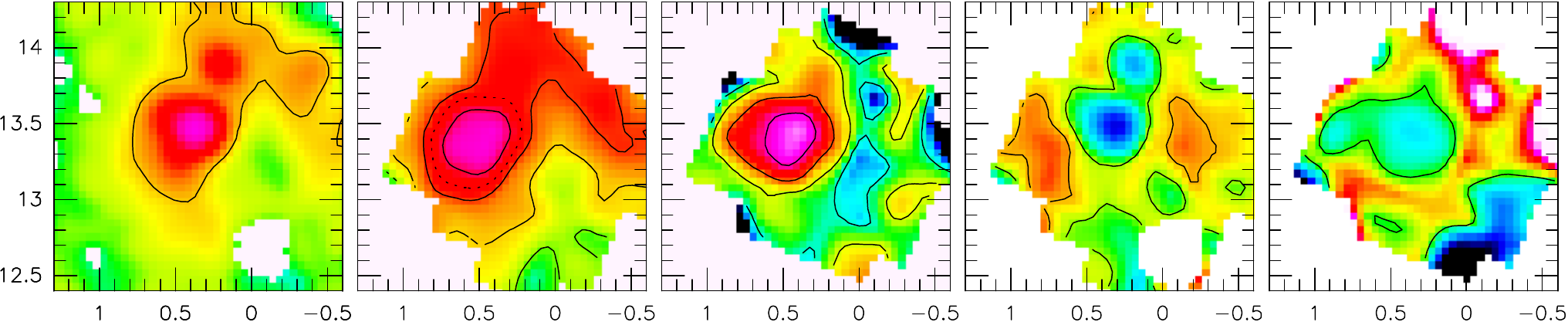}
 \caption{Three northern most regions of M\,33 covered with PACS. Top
   row: Northernmost region. Middle row: Northern region.  Bottom row:
   BCLMP\,691. From left to right: intensities of total infrared
   continuum (TIR), \CII, \OI(63$\mu$m), \CII/TIR, and
   \CII/\OI\ intensity ratios.  Intensities are in units of
   \ergintensity.  Offsets are in arcminutes relative to the nucleus,
   the same as in Fig.\,\ref{fig:strips-cen}. The TIR color wedge is
   the same as in Fig.\,\ref{fig:m33over}.  Contours of TIR, \CII,
   \OI, \CII/TIR, and \CII/\OI\ are the same as in
   Fig.\,\ref{fig:strips-cen}. The dashed \CII\ contour for BCLMP\,691
   corresponds to a constant intensity of $22.4\cdot 10^{-6}$
   \ergintensity\, derived from the best-fitting y-intercept in the
   correlation plot of $\log$\CII/TIR vs. $\log$TIR, for a slope of
   $-1$ (Fig.\,\ref{fig:cii-tir-bclmp691}).}
\label{fig:strips-north}
\end{figure*}

The region mapped with Herschel in \CII\ and \OI\ covers a radial
strip along the major axis of M\,33, which is, with a few gaps about
$35'$ long, and $1.5'$ wide (Figs.\ref{fig:m33over},
\ref{fig:strips-cen}, \ref{fig:strips-north}). The total area covered
is 38.5\,arcmin$^2$.
 Data were taken with PACS
 \citep{poglitsch.Waelkens.Geis.2010} on Herschel
 \citep{pilbratt.Riedinger.Passvogel.2010}.
\NII(122$\mu$m) data were also recorded but suffer from baseline
instabilities and are not discussed here. 
Most data were taken in unchopped mode with
an observation of an off-source position at the beginning and end of
each observation.  The off position for the two northern most
observations was at $1^{\rm h}35^{\rm m}28.87^{\rm s}$,
$+31^{\circ}46'1.92''$. For the other observations, an off position at
$1^{\rm h}30^{\rm m}20.90^{\rm s}$, $+30^{\circ}38'25.44''$ was used.
The only data taken in wavelength switching mode early-on in the
observation campaign are those of the \HII\ region BCLMP\,302 at
2\,kpc galacto-centric distance in the north of M\,33
\citep{mookerjea2011}. These data are included here for completeness.









The strip consists of 21 individual observations (ObsIds,
Table~\ref{tab:obssum}).  The footprint of the PACS array is
$47''\times47''$, resolved into $5\times5$ pixels. Each spatial pixel
(spaxel) has a size of $9.4''$. For each center position, a $3\times3$
raster was observed with a step-size of $24''$, resulting in a total
coverage of $95''\times95''$ per center position. The half power
beamwidths (HPBWs) for the \OI\ and \CII\ line observations are
$9.5''$ and $11.5''$, respectively.  Line widths in M\,33 are not
resolved: the instrumental spectral resolution is about 90\,\kms\ and
240\,\kms, respectively (PACS Observer's Manual V2.3), far larger than
line widths observed in M\,33 in \CII\ \citep{mookerjea2016,
  braine2012}, CO or \HI\ \citep{druard2014, gratier2017}.

  PACS data reduction of the unchopped scans was done using {\tt HIPE}
  version 15.0.1 \citep{ott2010}.  The data were calibrated using the
  PACS calibration tree version 78. The data pipeline was used,
  including a correction for transients (glitches) and including
  spectral flat fielding, to obtain level 2 data products for each of
  the individual 21 ObsIds.  Stepping through the pipeline and
  displaying intermediate results indicated that the ``transient
  correction'' is necessary to obtain good final data products.  At
  this point, the ``Off'' data products were kept separate from the
  ``On'' data products.  After running this initial pipeline, a script
  ``Spectoscopy: Mosaic multiple observations'' from the menu ``PACS
  Useful scripts'' was used to combine the individual ObsIds. This
  script was modified slightly to average all Off-positions together,
  subtract them from each On-position, and to create the final data
  cube.  The output is a single FITS file that contains the spectra at
  each spaxel of the combined map on a $3''$
  grid. Figure\,\ref{fig:spectra} shows a few exemplary spectra.

Python scripts were developed to derive line integrated intensities
and to estimate the baseline noise. Polynomials of up to third order
were fit to the spectra, masking the edges, which suffer from
increased noise and artifacts, and also the region around the expected
line position from \HI\ data \citep{warner1973}. The script determined
the best fitting polynomial, and subtracted it. Integrated line
intensities were derived by summing over a narrow wavelength range,
2.8 times the resolution, centered on the \HI\ velocities at each PACS
spaxel.

The baseline noise was estimated by calculating the standard deviation
(root mean square, rms) within the wavelength range used to fit the
baseline.  The spectral noise varies with spaxel.
%
%
The median baseline noise, that is, the statistical error, are
$(1.6\pm1.1)\cdot 10^{-7}$\,\ergintensity\, for \CII,
$(3\pm7)\cdot 10^{-7}$\,\ergintensity\, for \OI(63$\mu$m).
%
For each position the local 1~$\sigma$ uncertainty was estimated by
adding in quadrature the local noise levels and the relative
calibration errors of 15\% for \CII\ and \OI, respectively.
%
%
%



\begin{figure}[h!]
 \includegraphics[angle=0,width=0.48\textwidth]{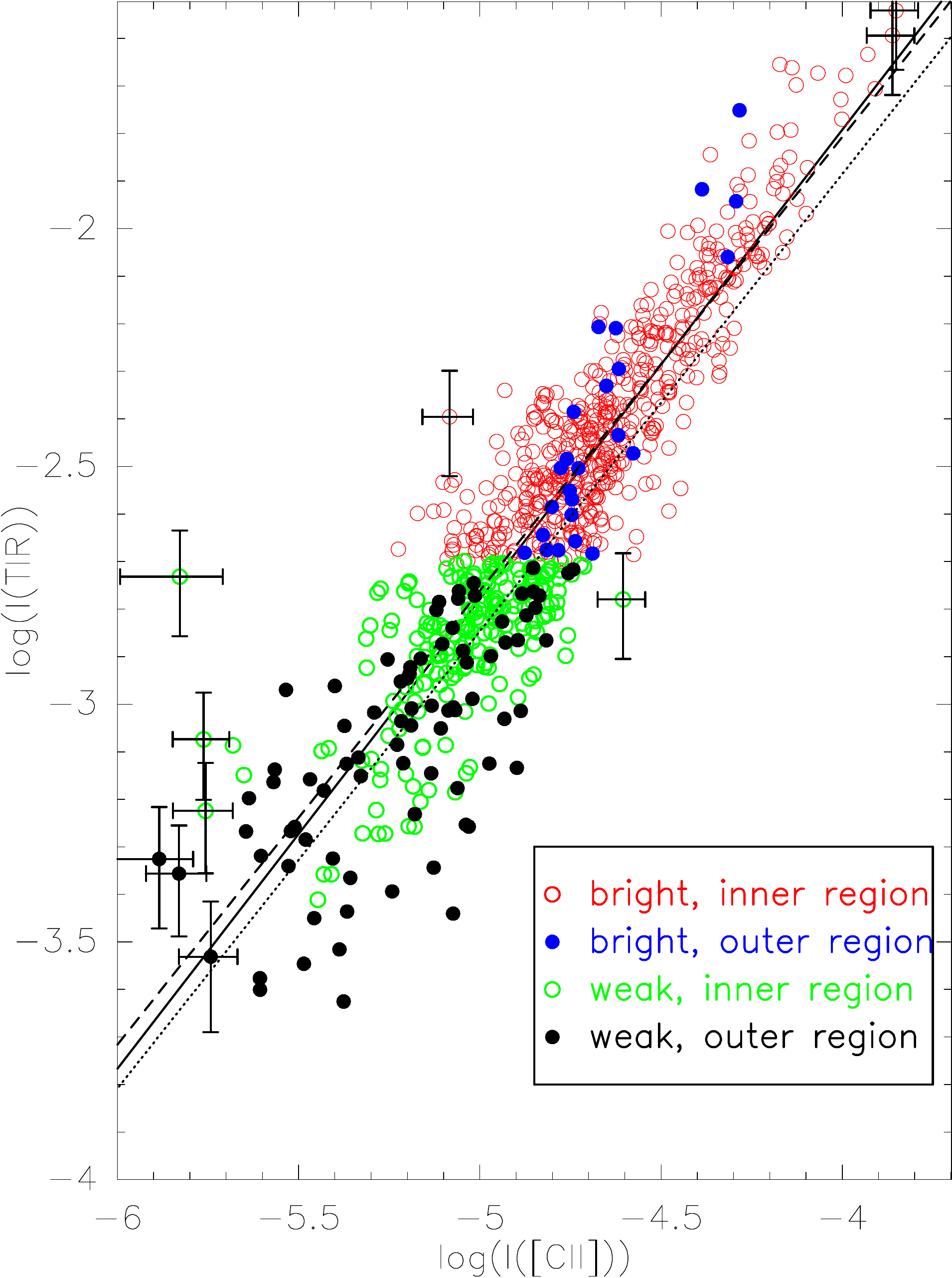}
 \caption{Total infrared continuum intensity (TIR) versus
   \CII\ intensities in M33. Intensities are given in \ergintensity.
   Each point corresponds to one position on a $12''$ grid at $12''$
   resolution, with an intensity 3 times above the local rms, for
   \CII, \OI, and TIR.  Typical $1\sigma$ errorbars are shown. Colors
   correspond to the four different regions of the inner and outer
   galaxy, and above and below a TIR threshold, as described in
   Sec.\,\ref{sec:ciitir}.  Straight lines delineate the results of
   unweighted linear least-squares fits to all data (drawn), the inner
   galaxy (dashed), and the outer galaxy (dotted)
   (Table\,\ref{tab:tir-cii}).  }
 \label{fig:tir-cii}
\end{figure}

\begin{figure*}[h!]
  \includegraphics[angle=0,width=0.9\textwidth]{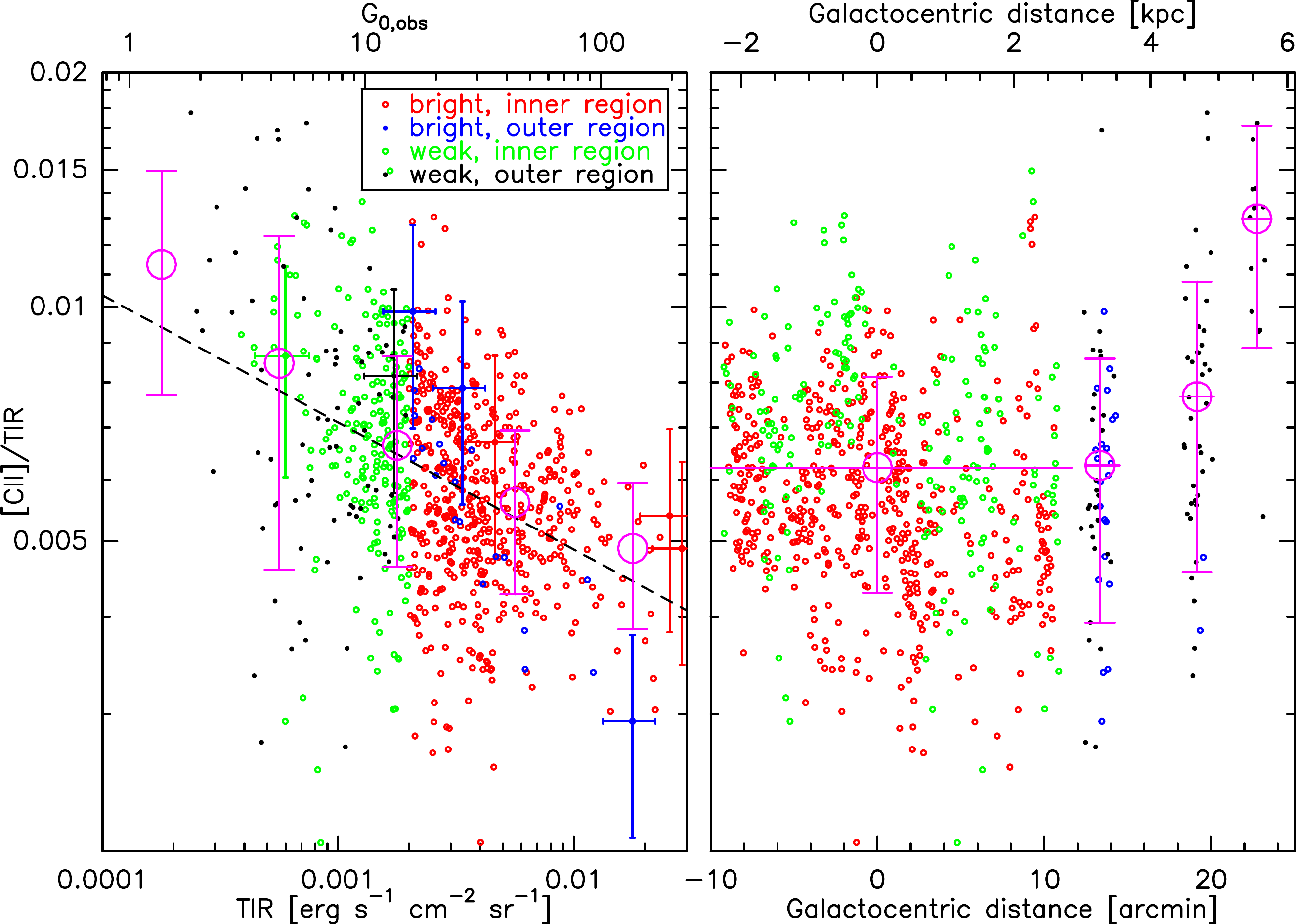}
  \caption{ ({\bf Left}) \CII/TIR intensity ratio versus TIR and
    corresponding FUV field \gobs\ in M\,33 (see equation\,\ref{eq:g0}
    in Sec.\,\ref{sec:ciitir}).  The large pink points and errorbars
    show binned ratios (Table\,\ref{tab:ciitirvstir}). Errorbars
    represent the standard deviation of the data. The dashed line
    shows the result of an unweighted least-squares fit to
    $\log$(\CII/TIR)$=b+m\times\log$TIR and all ratios, which gives
    $m=-0.16\pm0.10$ and $b=-2.64\pm0.26$, with correlation
    coefficient $r=-0.39$. For a few of the individual points
    $1\sigma$ errorbars are shown.  ({\bf Right}) \CII/TIR\ versus
    projected galacto-centric radius in M\,33. The large pink points
    and errorbars show ratios binned over selected ranges of radii
    (Table\,\ref{tab:ciitirvsradius}). }
 \label{fig:ciitir-tir-r} 
\end{figure*}




\subsection{Total infrared continuum and atomic hydrogen} 

The integrated intensity of the continuum emission between wavelengths of
1\,$\mu$m and 1\,mm, the TIR, has been estimated for each pixel from a
weighted sum of the MIPS/Spitzer 24~$\mu$m and PACS 70~$\mu$m,
100~$\mu$m, and 160~$\mu$m maps. Before combining these maps, all
images were convolved to a common resolution of $12''$ using the
dedicated kernels provided by \citet{aniano2011} and regridded to a
common reference frame with a pixel size of $3''$ projected onto the
\CII\ PACS positions. The PACS~70\,$\mu$m map \citep{boquien2015} was
rescaled to MIPS fluxes by multiplying with the MIPS scaling factor
($mc=1.03$) and dividing by the color correction factor ($cc=0.98$)
which were interpolated from Table\,3 of the PACS report by
\citet{mueller2011} assuming a modified blackbody with an emissivity
index of $\beta=1.5$ and an average dust temperature of 35\,K
\citep{tabatabaei2014}: $S_{70}^{\rm{MIPS}} = mc/cc\cdot
S_{70}^{\rm{PACS}}$.  Next, the TIR was calculated using weighting
factors from Equation\,1 and Table\,1 of \citet{boquien2011}, and
using all data larger than zero:

\begin{eqnarray}
  \label{eq:tir}
  \log S_{\rm{TIR}} & = & 0.220 \cdot \log S_{24} + 
                          0.202 \cdot \log S_{70}^{\rm{MIPS}} + \nonumber \\ 
                    &   & 0.243 \cdot \log S_{100} + 
                          0.311 \cdot \log S_{160} +
                          1.361.
\end{eqnarray}

In a similar approach, \citet{galametz2013} used spatially resolved
Herschel maps of the {\tt KINGFISH} sample of nearby galaxies to
derive weighting factors to estimate the TIR. \citet[][cf. their
  Fig.\,6]{galametz2013} find that the determined calibration
coefficients are very similar to those measured in M\,33 by
\citet{boquien2011} using Spitzer and Herschel maps.

The median statistical error of the TIR in M\,33 as measured in
emission free areas, is $(2.0\pm1.0)\cdot 10^{-5}$\,\ergintensity. For
each position, the local 1~$\sigma$ uncertainty was estimated by
adding in quadrature the statistical error and the relative
calibration error of 25\%, assuming 20\% error on the 160$\,\mu$m
data, and $\sim10$\% error on the 100, 70, and 24$\,\mu$m data,
respectively \citep{boquien2011}.

The TIR map (Fig.\,\ref{fig:m33over}) resulting from
Eq.\,\ref{eq:tir}, shows the flocculent spiral structure of M\,33, a
number of prominent giant \HII\ regions, the arm and interarm regions,
together with the drop of intensities with increasing galacto-centric
radius.


To estimate the contribution of the cold neutral medium to the
\CII\ emission, we used a VLA map of M\,33 in the atomic hydrogen
21\,cm line at $12''$ resolution \citep{gratier2010}, matching the
resolutions of the other data.

\section{Analysis}

\subsection{\CII\ emission and the total infrared continuum}
\label{sec:ciitir}


The maps of TIR and \CII\ (Figs.\,\ref{fig:m33over},
\ref{fig:strips-cen}, \ref{fig:strips-north}) show a striking
similarity. Both tracers vary by more than two orders of magnitude in
intensity. They are bright in the inner star forming arms and
gradually fainter in the outskirts at several kiloparsec radial distance. A
correlation plot of TIR against \CII\ shows their tight correlation
(Fig.\,\ref{fig:tir-cii}). The results of linear
  least-squares fits to $\log\,I$(TIR) against $\log\,I$(\CII) are shown
  in Fig.\,\ref{fig:tir-cii} and Table\,\ref{tab:tir-cii}.
The scatter is larger at fainter intensities and larger than the
statistical errors of the individual positions. To gain more insight,
we have subdivided the observed positions between those of TIR
intensities above and below a given value, tracing the dense, warm,
star-forming spiral arms on the one hand, and the diffuse inter-arm
dust on the other hand. A second subdivision distinguishes between the
inner and outer galaxy. The TIR threshold and the radial boundary have
no specific physical meaning and were selected to emphasize
differences between the four regions. We chose a TIR threshold of
$2\cdot 10^{-3}$\,\ergintensity\, (shown as contour in
Figs.\,\ref{fig:strips-cen} and \ref{fig:strips-north}) and a radial
boundary of 2.93\,kpc ($12'$) to separate the inner and outer galaxy
\citep[cf. ][]{verley2009}.
%
%
%
The fit slopes and intercepts of the inner and outer regions of the
galaxy do not differ within the errors (Table\,\ref{tab:tir-cii}). We
find indications for a steepening of the slope for the TIR-bright
outer regions (Fig.\,\ref{fig:tir-cii}).

As the dust grains emitting the observed TIR emission are heated by
the stellar radiation field, the TIR can be used to estimate the FUV
energy flux \gobs,
%
%
which is given in units of the average radiation field in the solar
neighborhood, the Habing field of
$1.6\cdot 10^{-3}$\,erg\,cm$^{-2}$\,s$^{-1}$ \citep{habing1968}:

\begin{equation}
  G_{0, \rm{obs}} = C_1 C_2 \,\frac{\rm{TIR}}
  {1.6\cdot 10^{-3}/(4\pi)}
\label{eq:g0}
\end{equation}

 with TIR in units of \ergintensity. The factor of $C_1=0.5$
 approximately takes into account the absorption of visible photons by
 grains \citep{kaufman1999, tielens1985}.

The factor $C_2$ corrects for the fraction of FUV photons leaking out
of the galaxy. To estimate the FUV attenuation in M\,33,
\citet{boquien2015} combined GALEX FUV maps with 24$\mu$m maps,
following \citet{kennicutt2009}. The typical attenuation in M\,33 is
around 0.6\,mag in the FUV band, consistent within the scatter with
0.53\,mag previously found by \citet{verley2009}. In the low
metallicity environment of M\,33, about half of the FUV photons are
hence on average absorbed by dust and reradiated in the TIR, while the
other half escapes, that is, $C_2=2$.


In M\,33 at 50\,pc resolution, the estimated FUV field \gobs\ ranges
between $\sim2$ and 200 (Fig.\,\ref{fig:ciitir-tir-r}). The selected
TIR threshold of $2\cdot 10^{-3}$\,\ergintensity\, corresponds to a FUV
field of \gobs$=15$. Further below, we used the observed
(\CII$+$\OI)/TIR and \CII/\OI\ ratios together with PDR models of
given density and FUV field to improve on this estimate.

    
The average \CII/TIR ratio in M\,33 is ($0.64\pm0.21$)\%
(Table\,\ref{tab:ciitirvstir}). Figure\,\ref{fig:ciitir-tir-r} (Left)
shows the data at the individual positions, together with unweighted
binned averages. For low TIR values the \CII/TIR ratio shows a large
scatter with a high averaged value of ($1.1\pm0.4$)\% at
$\log$(TIR)=$-3.75$, with the TIR intensities in units of
\ergintensity. The binned ratios drop smoothly with TIR, reaching
($0.5\pm0.1$)\% at high TIR values of $\log$(TIR)=$-1.75$.  The binned
averages are consistent with the results of a linear least-squares fit
(Fig.\,\ref{fig:ciitir-tir-r}).

To explore the \CII/TIR variation further, we study its variation with
galacto-centric distance (Fig.\,\ref{fig:ciitir-tir-r}, right). The
\CII/TIR ratio stays almost constant within a galacto-centric radius
of $\sim4\,$kpc \citep[cf.][]{verley2009}. The two northern regions
show rising \CII/TIR ratios, as had already been seen in the ISO/LWS
cut (K2013).  The northernmost region shows a mean ratio that is a
factor 2 higher than in the inner galaxy, $1.3\%$, with an increased
scatter of $0.4\%$ (Table\,\ref{tab:ciitirvsradius}). The
corresponding map (Fig.\,\ref{fig:strips-north}) shows a small region
of low \CII/TIR ratios, where TIR is just above the threshold of
$2\cdot 10^{-3}$\,\ergintensity, but surrounded by an extended region of
high ratios $>1\%$ where TIR is below the threshold and \CII\ is also
weak.

The ISO/LWS cut along the major axis of M\,33 (K2013) shows a flat
\CII/FIR\footnote{K2013 discussed the far-infrared emission (FIR),
  integrated between 42.5 and 122.5$\mu$m.}  distribution in the inner
galaxy, and an abrupt rise of the average ratio in the northern and
southern outskirts, at galacto-centric distances beyond $\sim4.5$\,kpc
out to $\sim7.5$\,kpc. Using the FIR/TIR correction factors listed for
each position in Table\,A.1 of K2013, the resulting \CII/TIR values
are consistent within the $1\sigma$ scatter with the ratios determined
here. The Herschel/PACS data presented here improve on this work due
to their high sensitivities at much better angular resolution, mapping
the GMCs also perpendicular to the major axis of M\,33.

\begin{table}
  \caption{Results of unweighted linear least-squares fits to
    $\log$TIR=$b+m\times\log$\CII\ (Fig.\,\ref{fig:tir-cii}) with TIR
    and \CII\ in units of \,\ergintensity. The Pearson correlation
    coefficient $r$ serves as a measure of the linear correlation
    between TIR and \CII.}
\label{tab:tir-cii}
\centering
\begin{tabular}{rlrrr}
\hline\hline
 Region        & $m\pm\sigma$  & $b\pm\sigma$ & $r$ \\
\hline 
      all      &  $0.99\pm0.11$ & $2.15\pm0.52$ & 0.91 \\
         inner &  $0.95\pm0.13$ & $2.01\pm0.60$ & 0.90 \\ 
         outer &  $0.96\pm0.29$ & $1.96\pm1.50$ & 0.87 \\ 
%
%
\hline
\end{tabular}
\end{table}


\begin{table}
  \caption{Binned \CII/TIR ratios shown in
    Figure\,\ref{fig:ciitir-tir-r}. The width of the TIR bins is
    0.5\,dex. Errors of the individual points are ignored.}
\label{tab:ciitirvstir}
\centering
\begin{tabular}{rr}
\hline\hline
 $\log$ TIR & \CII/TIR  \\
& \% \\
\hline
 all & $0.64\pm0.23$ \\ 
 \hline
 $-3.75$ & 1.13 $\pm$ 0.36 \\ 
 $-3.25$ & 0.85 $\pm$ 0.39 \\ 
 $-2.75$ & 0.66 $\pm$ 0.20 \\ 
 $-2.25$ & 0.56 $\pm$ 0.13 \\ 
 $-1.75$ & 0.49 $\pm$ 0.10 \\ 
\hline
%
\end{tabular}
\end{table}

\begin{table}
  \caption{\CII/TIR ratios averaged over different galacto-centric
    distances (cf. Figure\,\ref{fig:ciitir-tir-r}). Errors of the individual points are ignored.}
  \label{tab:ciitirvsradius}
\centering
\begin{tabular}{rrr}
\hline\hline
 \multicolumn{2}{c}{radial range} & \CII/TIR \\
 arcsec           & kpc           & \%     \\
\hline
 $-700..700$ & $-2.8..2.8$ & $0.62\pm0.19$ \\
 $750..870$  & $3.1..3.5$  & $0.63\pm0.23$ \\
 $1090..1210$ & $4.4..4.9$ & $0.77\pm0.31$ \\
 $1320..1410$ & $5.4..5.7$ & $1.30\pm0.41$ \\
%
 \hline
\end{tabular}
\end{table}

\begin{table}
  \caption{Results of unweighted linear least squares fits to
    $\log$\OI$=b+m\times\log$\CII.  (Fig.\,\ref{fig:oivscii}).}
  \label{tab:oicii}
\centering
\begin{tabular}{rlrr}
\hline\hline
 Region & $m\pm\sigma$ & $b\pm\sigma$ & $r$ \\
 \hline
 bright, inner & $0.84\pm0.01$ & $-1.43\pm0.86$ & 0.74 \\
 bright, outer & $1.49\pm0.25$ & $1.65\pm5.65$ & 0.75 \\
 \hline
\end{tabular}
\end{table}

\begin{figure}[h!]
 \includegraphics[angle=0,width=0.48\textwidth]{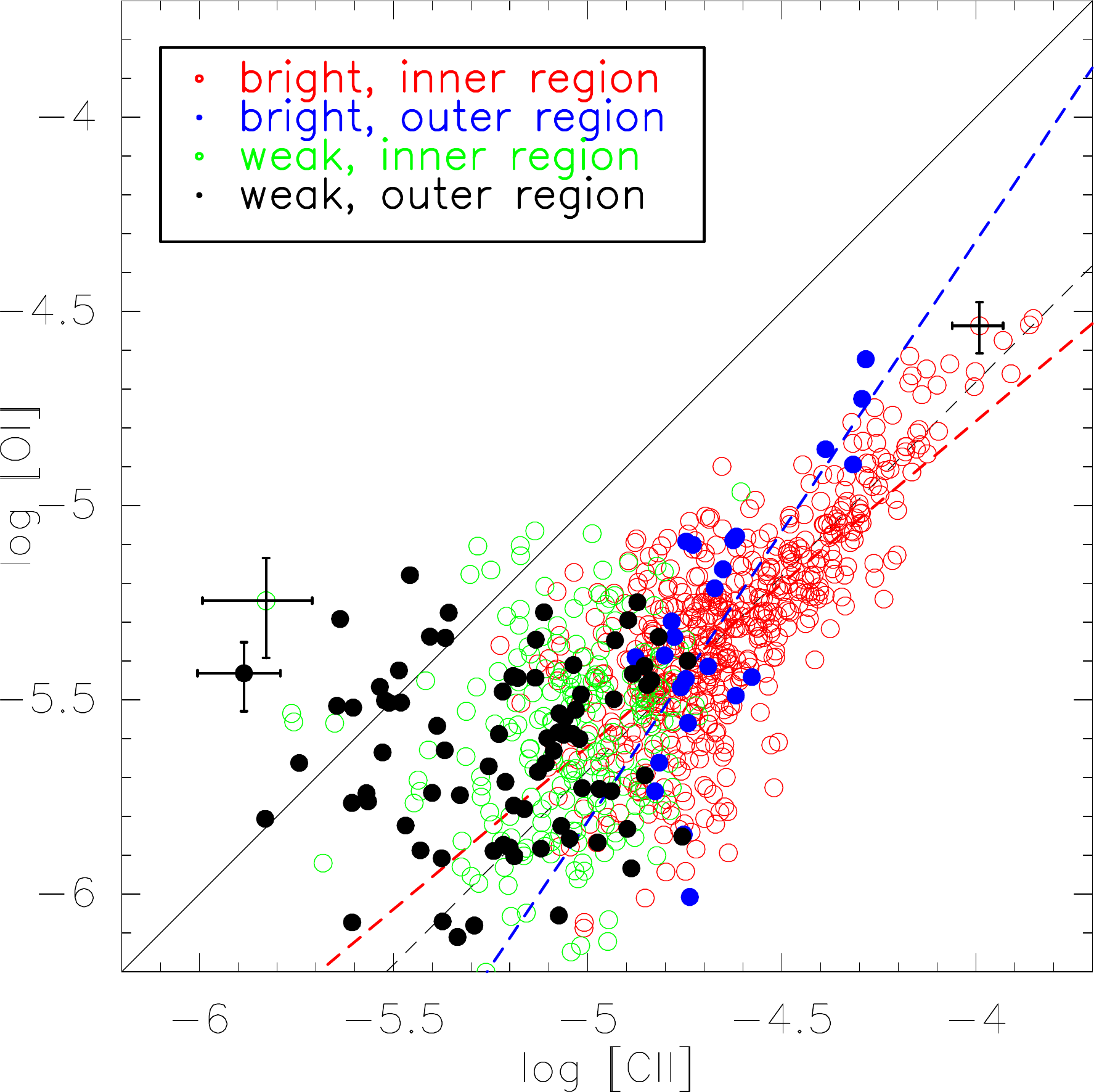}
 \caption{Observed intensities of \OI(63$\mu$m)\ and \CII\ in M\,33
   with representative $1\sigma$ errorbars. The straight black line
   corresponds to a ratio of 1.  The dashed black line corresponds to
   the average \CII/\OI\ ratio (Table\,\ref{tab:cii-oi-tir}). The red
   and blue dashed lines are the results of unweighted least squares
   fits to the TIR-bright inner and outer regions, respectively
   (Table\,\ref{tab:oicii}).  }
 \label{fig:oivscii} 
\end{figure}


\begin{figure*} 
  \includegraphics[angle=0,width=0.49\textwidth]{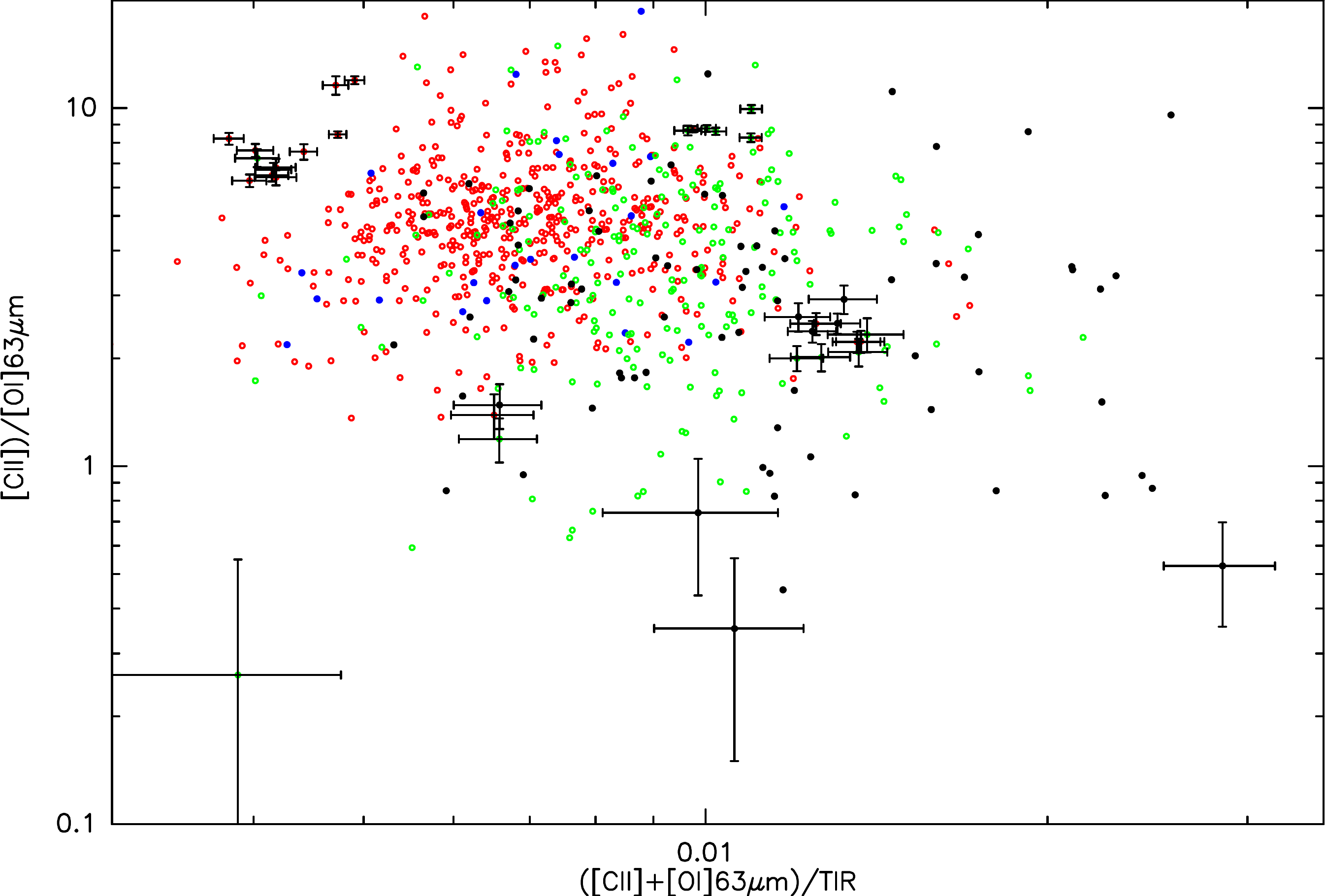}
  \includegraphics[angle=0,width=0.49\textwidth]{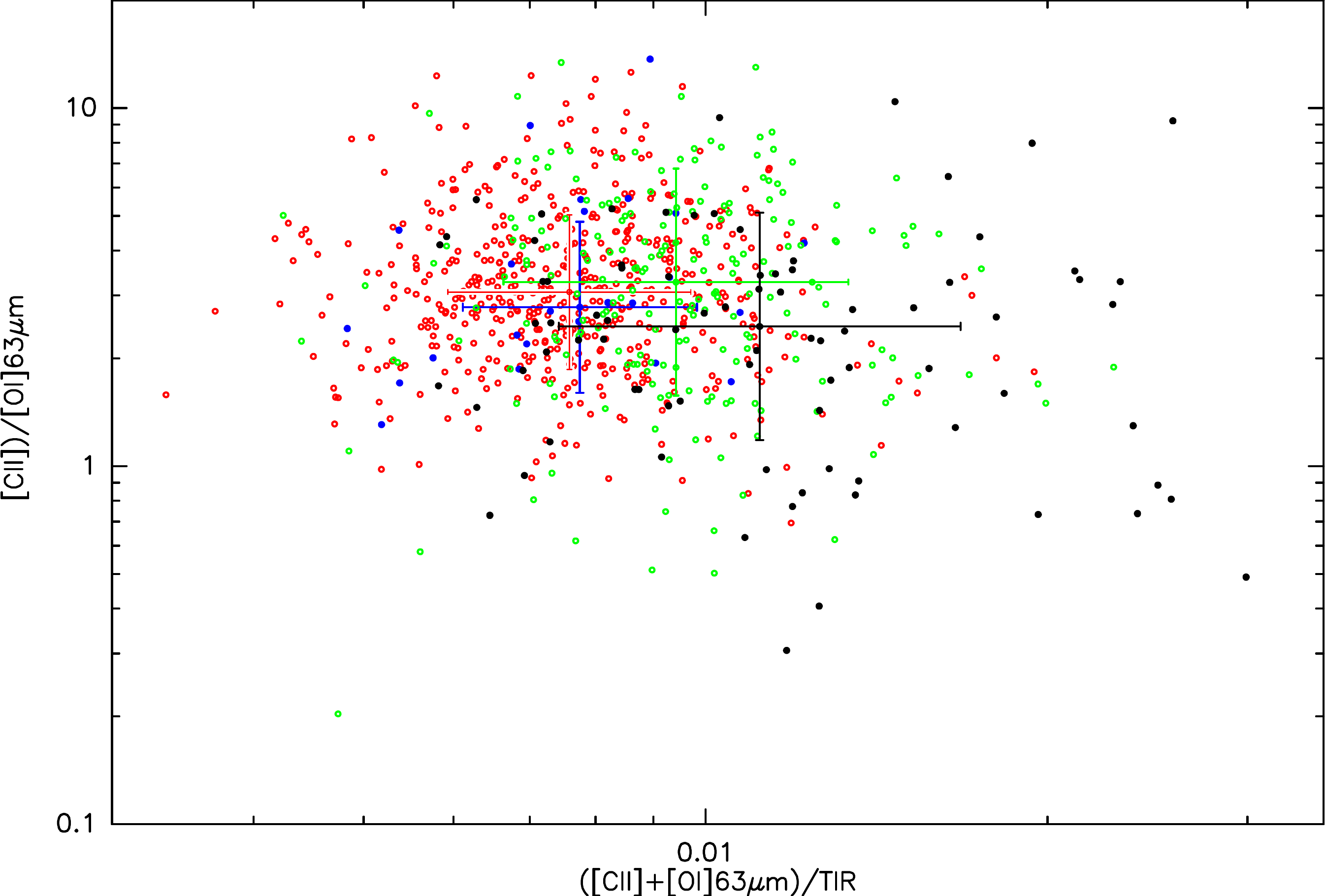}
  \caption{ \CII/\OI(63$\mu$m) ratio plotted against the
    (\CII+\OI($63\,\mu$m)/TIR ratio for M\,33. \CII, \OI, and TIR
    intensities are from positions with intensities above the local
    $3\sigma$ value. Colors correspond to the four different regions
    of the inner and outer galaxy, and above and below a TIR
    threshold, as described in Sec.\,\ref{sec:ciitir}. {\bf Left:}
    Observed ratios with typical 1$\sigma$ errors. {\bf Right:} Ratios
    with corrected \OI\ intensities, as described in
    Sec.\,\ref{sec:oicorr}. In addition, median and standard
    deviations are shown for the four regions (errorbars in color)
    (cf. Table\,\ref{tab:cii-oi-tir}).}
  \label{fig:ciioitir-observed}
\end{figure*}

\begin{figure*} 
  \includegraphics[angle=0,width=12cm]{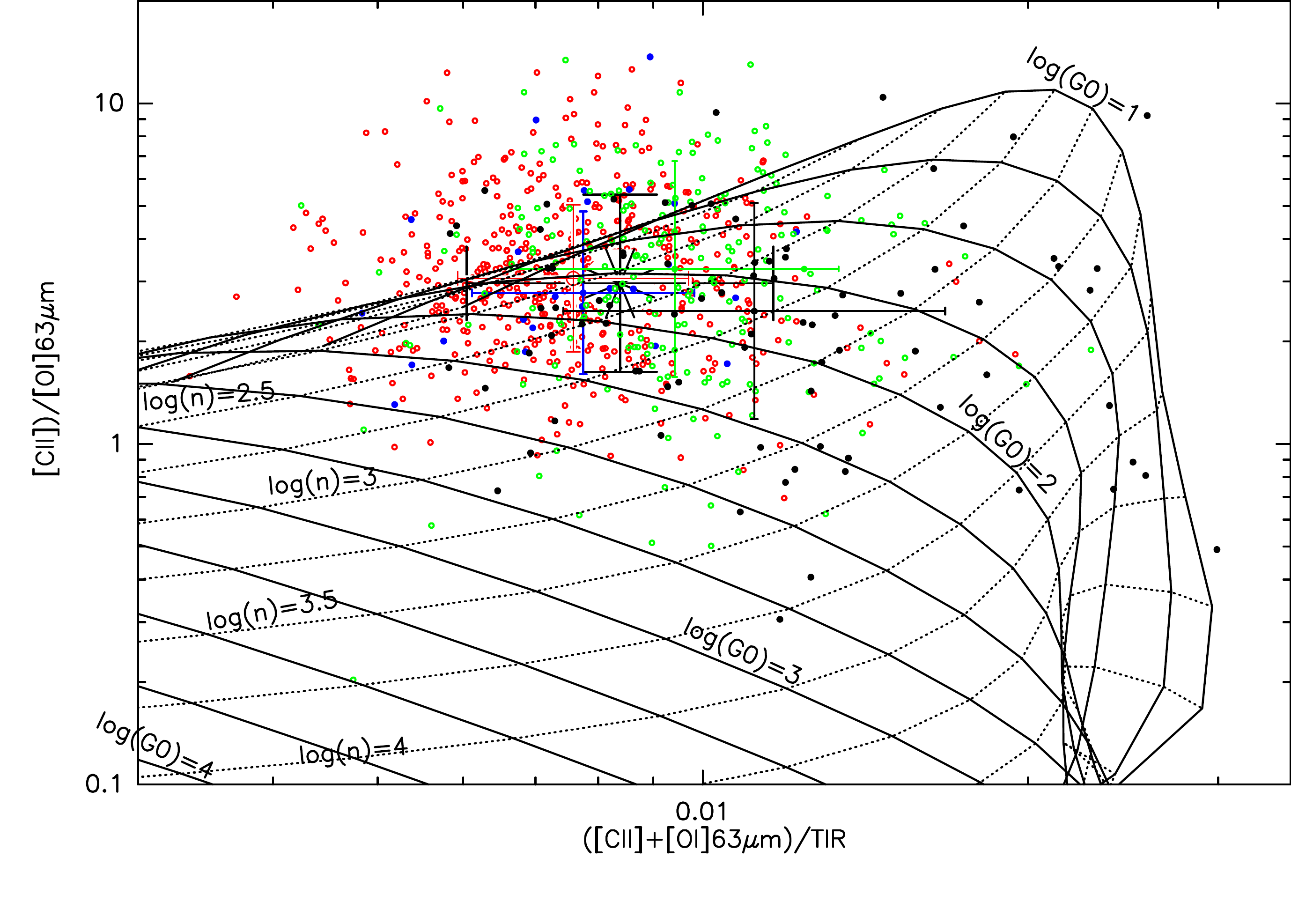} \\
  \sidecaption
  \includegraphics[angle=0,width=12cm]{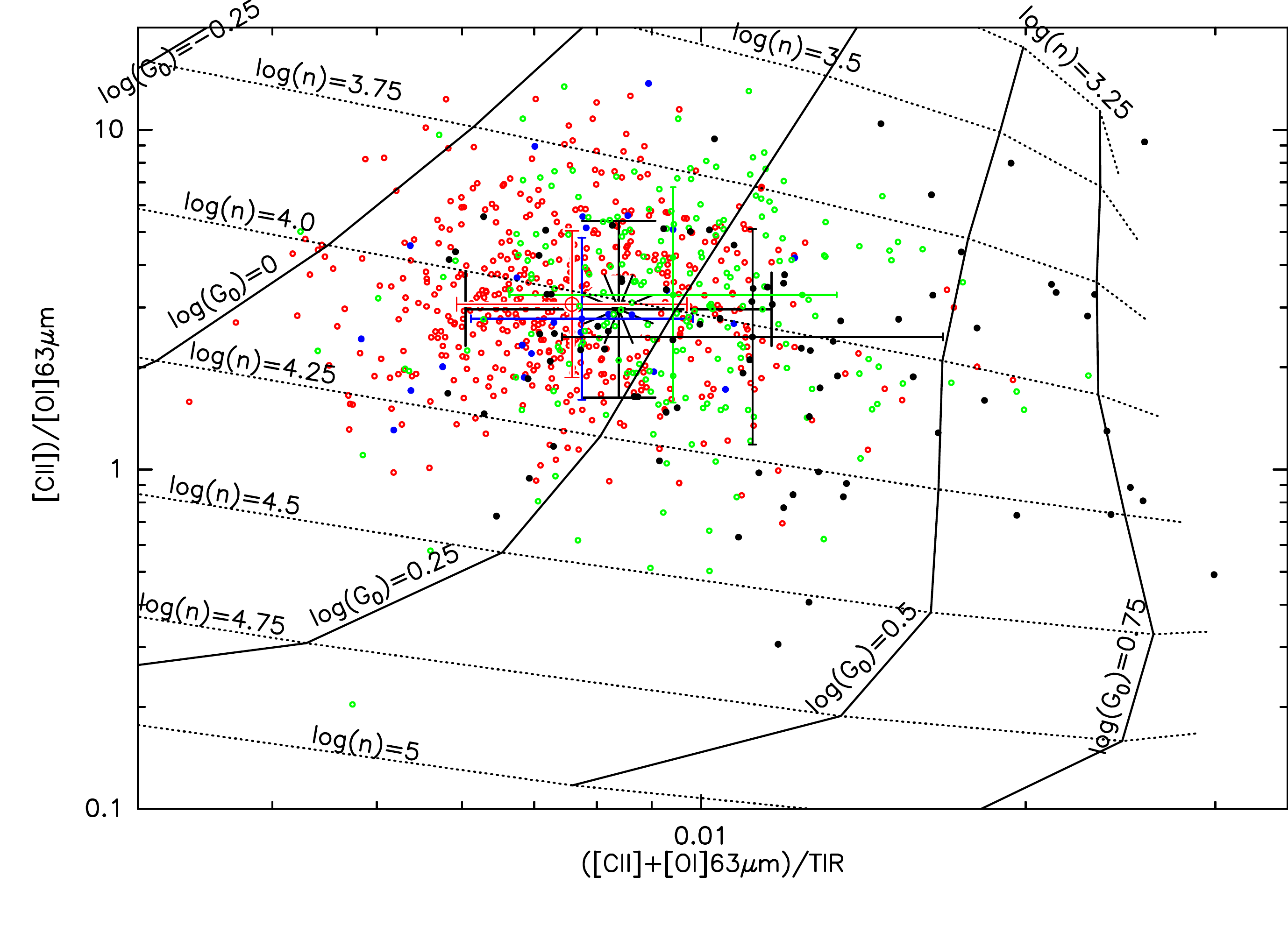} 
 \caption{Diagnostic diagram of the \CII/\OI(63$\mu$m) ratio plotted
   against the (\CII+\OI($63\,\mu$m)/TIR ratio for M\,33, using the
   corrected \OI\ intensities (see Sec.\,\ref{sec:oicorr}). Median and
   standard deviation of the four regions are shown in color, and of
   all data (star and large errorbars). Colors correspond to the four
   different regions as described in Sec.\,\ref{sec:ciitir}: the TIR
   bright, inner region (red), the bright, outer region (blue), the
   weak, inner region (green), and the weak, outer region (black).
   Superimposed is a grid of constant hydrogen nuclei density $n$
   (dashed contours) and FUV field strength $G_0$ (solid contours)
   from PDR models \citep{kaufman1999, kaufman2006}. We note that for
   many ratios there are two solutions for $n$, $G_0$. {\bf Upper
     panel:} Solutions with moderate $n$, $G_0$. {\bf Lower panel:}
   Low-FUV solutions. }
 \label{fig:ciioi} 
\end{figure*}


\begin{table*}
  \caption{Median \CII/\OI\ and (\CII+\OI)/TIR ratios and rms values
    as observed and after correction for \OI\ self absorption.
    Solutions of the PDR models of \citet{pound2008,kaufman2006} are
    shown for the median ratio of all positions and for position \#1
    with one of the highest observed \CII\ intensities $1.2\cdot
    10^{-4}$\,\ergintensity. }
\label{tab:cii-oi-tir}
\centering
\begin{tabular}{rrrrrrr}
\hline\hline
Region & \CII/\OI & (\CII+\OI)/TIR & \multicolumn{4}{c}{Solutions of PDR models} \\
&          &                &
  \multicolumn{2}{c}{Moderate solution} & \multicolumn{2}{c}{low-FUV solution} \\
       &          &                & $n$ & $G_0$ & $n$ & $G_0$ \\
       &          &     \%         & [cm$^{-3}$] & & [cm$^{-3}$]     \\
 \hline
               & \multicolumn{2}{c}{Observed ratios} \\
 all           & $4.51\pm2.58$ & $0.75\pm0.31$ &  60 & 20 & $10^4$ & 1 \\
 bright, inner & $4.79\pm2.43$ & $0.68\pm0.19$ &  60 & 20 & $10^4$ & 1 \\
 bright, outer & $3.71\pm3.68$ & $0.72\pm0.19$ &  30 & 20 & $10^4$ & 1 \\
 weak, inner   & $3.83\pm2.54$ & $0.91\pm0.27$ & 100 & 30 & $10^4$ & 1 \\
 weak, outer   & $3.01\pm2.36$ & $1.07\pm0.54$ & 300 & 60 & $10^4$ & 1 \\
 \hline
 & \multicolumn{2}{c}{Ratios with corrected \OI} \\
 all           & $3.0\pm2.1$   & $0.82\pm0.53$ & $200$ & 60 & $10^4$ & 1.5 \\ 
 bright, inner & $3.06\pm1.92$ & $0.76\pm0.2$  & $200$ & 60 & $10^4$ & 1.5 \\ 
 bright, outer & $2.78\pm2.69$ & $0.77\pm0.18$ & $200$ & 70 & $10^4$ & 1.5 \\ 
 weak, inner   & $3.27\pm2.28$ & $0.94\pm0.89$ & $300$ & 60 & $10^4$ & 1.5 \\ 
 weak, outer   & $2.46\pm2.02$ & $1.12\pm0.55$ & $600$ & 80 & $10^4$ & 2.0 \\ 
 \hline
 \#1           & $4.59\pm0.97$ & $0.76\pm0.25$ & $900$ & $20$ & $5\cdot 10^3$ & 3.0 \\ 
\hline
\end{tabular}
\end{table*}

\subsection{Emission of \CII, \OI\ 63$\mu$m, and the TIR}

The critical densities of the \CII(158$\mu$m) and \OI(63$\mu$m) lines
for excitation by collisions with H are $3\cdot 10^3$\,\cmcub\ and
$5\cdot 10^5$\,\cmcub, respectively, with upper energy levels $E/$k$_B$ of
92\,K and 228\,K, respectively \citep{kaufman1999}. In contrast to the
\CII\ line, the \OI(63$\mu$m) line is expected to be excited almost
exlusively in the dense, warm interface regions of UV illuminated
molecular clouds. The distribution of these two emission lines
reflects these strong differences in excitation requirements. The
intensities of \CII\ and \OI\ emission are well correlated near the
peaks and ridges of the spiral arms of M\,33, as is seen in the maps
of the central region and of BCLMP\,691 (Figs.\,\ref{fig:strips-cen},
\ref{fig:strips-north}) and in the correlation plot
(Fig.\,\ref{fig:oivscii}, cf. Table\,\ref{tab:oicii}). At lower
intensity levels of \CII\ and \OI, in the more diffuse inter-arm
regions, the correlation between the two gas tracers is weak. This is
seen most prominently in the maps of the two northern most regions
(Fig.\,\ref{fig:strips-north}).

As expected, \CII\ is stronger than \OI\ at most positions. The
observed \CII/\OI($63\mu$m) ratio ranges between $\sim0.3$ and 20 with
a median of $4.5\pm2.6$. The observed (\CII+\OI($63\,\mu$m)/TIR
ratios vary between 0.3 and 2.9\% with a median of $0.8\pm0.3$\%
(Fig.\,\ref{fig:ciioitir-observed}).  Table\,\ref{tab:cii-oi-tir}
lists the average values and the scatter for the four regions defined
in Section\,\ref{sec:ciitir}. The TIR bright positions all show ratios
of \CII/\OI$>1$ and (\CII+\OI)/TIR$<$2\%, while the positions where
\OI\ intensities exceed those of \CII, and (\CII+\OI)/TIR$>2$\% all
lie in the TIR weak regions.

PDR models may allow one to estimate the local excitation conditions,
the local densities and the local FUV fields. In order to compare the
observations with the predictions of PDR models, we need to consider
other components of the ISM that may contribute to the emission. A
part of the \CII\ emission may stem from the cold neutral medium (CNM)
or from a diffuse ionized medium. The \OI(63$\,\mu$m) line may become
optically thick and be affected by foreground absorption. Below, we
discuss these possibilities.

\begin{table}
    \caption{Assumptions to estimate the fraction of \CII\ emission
      from the cold neutral medium (CNM). $N$(\HI,PDR) is the
      estimated column density of atomic hydrogen at the surfaces of
      the PDRs. $X$(C$^+$) is the fractional abundance of C$^+$,
      $n_{\rm CNM}$ and $T_{\rm CNM}$ are density and temperature of
      the CNM.}
    \label{tab:ciicnm}
    \centering
    \begin{tabular}{rl}
      \hline\hline
      $N$(\HI,PDR) & $3.25\cdot 10^{20}$\,cm$^{-2}$ \\
      $X$(C$^+$)   & $5.9\cdot 10^{-5}$ \\
      $n_{\rm CNM}$ & 100\,cm$^{-3}$ \\
      $T_{\rm CNM}$ & 80\,K \\ 
      \hline
    \end{tabular}
\end{table}

\subsubsection{\CII\ from the cold neutral medium}

To estimate the contribution of the cold neutral medium (CNM) to the
observed \CII\ emission, we used \HI\ VLA data, at all positions at
which \CII\ has been detected, and at almost the same angular
resolution.
Following the approach of K2013, we first corrected the \HI\ line
intensities for the contribution from the surfaces of PDRs assuming a
typical $G_0/n$ ratio of 10$^{-3}$. Next, the \CII\ emission from the
remaining neutral gas, the CNM, was estimated assuming a given
fractional abundance of C$^+$/H in the low-metallicity environment of
M\,33, optically thin \HI\ and \CII\ emission, and a density and
temperature which are typical for diffuse atomic clouds
(Table\,\ref{tab:ciicnm}).  While some regions at larger
galacto-centric distances show enhanced CNM fractions, as already seen
by K2013, the average CNM contribution is only $\sim10$\%
(Fig.\,\ref{fig:iciicnmfrac}), less than the estimated
\CII\ calibration error. We do not subtract this small contribution
from the CNM from the observed \CII\ intensities to estimate the
\CII\ emission from PDRs.



\subsubsection{\CII\ from the ionized medium}

 Another part of the \CII\ emission may not arise from the neutral gas
 of PDRs modeled by \citet{kaufman2006}, but from the diffuse,
 ionized gas. Observations of the \NII(122$\mu$m, 205$\mu$m) lines
 would allow us to determine electron densities and the importance of
 the ionized gas \citep{oberst2006, croxall2012, croxall2017}.
 \citet{parkin.2013, parkin.2014} studied maps of \CII, \OI, and TIR
 emission in M~51 and Centaurus~A. PDR models better fit the observed
 \CII/\OI\ and \CII$+$\OI/TIR ratios after correcting for the
 fraction of \CII\ emission originating from the ionized medium using
 \NII(122$\mu$m, 205$\mu$m). They estimate that 70-80\% of \CII\ arise
 in ionized gas of the nucleus and center of M\,51, 50\% in its arm
 and interarm regions. In Centaurus~A the ionized fraction is
 10-20\%. \citet{hughes2015} use the \NII(205\,$\mu$m) line to
 estimate fractions of $\sim16-64\%$ in \object{NGC\,891}. They also
 explore the use of an empirical relation between \NII\ and 24\,$\mu$m
 emission, to construct a more complete map of the \CII\ fraction from
 the ionized gas. For M\,33, we do not attempt to correct the
 \CII\ intensities for the contribution from the ionized gas as
 observations of the \NII\ lines are missing.
%

\subsubsection{TIR}

A fraction of the TIR emission may stem from non-PDR phases of the ISM
like the CNM.  We did not try to correct the TIR emission for these
contributions before using the PDR models. However, the correlation
between TIR and \HI\ intensities is poor in M\,33
(Fig.\,\ref{fig:tir-hi}). An unweighted linear least-squares fit gives
a correlation coefficient of $r=0.45$, which is much lower than for
the TIR-\CII\ relation. This indicates that most of the TIR emission
stems from PDR regions, as the CNM contribution to the \CII\ emission
is also low.
 
\subsubsection{Self-absorbed \OI\ (63$\,\mu$m) emission}
\label{sec:oicorr}

The \OI(63$\mu$m) line is expected to have a higher opacity than the
\CII\ line and hence may be affected by self-absorption caused by cold
foreground clouds of atomic oxygen, as has been seen in velocity
resolved spectra and with narrow beams toward bright background
sources in the Milky Way \citep{schneider2018, gerin2015, leurini2015,
  karska2014, lis2001, timmermann1996}.  Spectra of the \OI(63$\mu$m)
line taken toward ULIRGs are often heavily self-absorbed
\citep{rosenberg2015}.  \citet{israel2017} also used PACS/Herschel to
observe the center of \object{Centaurus~A}. Using PDR models allowed
them to compare the observed \OI(63\,$\mu$m) intensity with the
intensity predicted from modeling the emission of other FIR lines.
In the circumnuclear disk \citet{israel2017} find high optical depths
of the \OI\ line of $1.0-1.5$. While these PDR models take into
account optical depth effects of the spectral lines emerging from the
simulated slab of gas, they do not consider possible absorption by
foreground gas of different excitation conditions.


The intrinsic line widths observed in M\,33 in \CII, CO and \HI\ are
far smaller than the velocity resolution of PACS/Herschel
($\sim90$\,\kms).  Using HIFI/Herschel {\tt HerM33es} observations,
\citet{mookerjea2016} and \citet{braine2012} find full-widths at
half-maximum (FWHM) of \CII\ of 7 to 17\,\kms\ in three regions along
the major axis of M\,33: BCLMP\,302, BCLMP\,691, and the nucleus.
\cite{druard2014} and \cite{gratier2017} studied the variation of
FWHMs derived from maps of the complete galaxy in CO 2--1 taken with
the IRAM 30m telescope and \HI\ taken with the VLA.  The line widths
averaged in radial bins of 1\,kpc drop with distance, out to 7.5\,kpc:
from about 8 to 5 \kms\ for CO, and from about 17 to 12\,\kms\ for
\HI. As the \OI(63$\mu$m) line is expected to arise from the dense,
warm cloud interfaces, its linewidths should be similar to those of
CO, and smaller than those of \CII, which may also trace other, more
diffuse phases.

Assuming low densities in the cold, foreground layers of gas, the
majority of the oxygen atoms are in their ground state, allowing to
derive an upper limit of the \OI\ opacities. The column density of
atomic oxygen can then be approximated in terms of the opacity
$\tau_0$ at the center velocity of the $J=1-2$ \OI\ 63$\,\mu$m line by

  \begin{equation}
    N(\rm{OI}) = 2\cdot 10^{17} \, \tau_0 \, \Delta v_{\rm{FWHM}}
    \label{eq:oisimple}
  \end{equation}

  \noindent in cm$^{-2}$, with the FWHM line width in
  \kms\ \citep{vastel2002, liseau2006}. For a typical line width of
  7\,\kms\ (see above) and an average oxygen abundance of $4\cdot 10^{-4}$
  in M\,33 (see below), the line center opacity exceeds 1 for
  $N($\OI$)=1.4\cdot 10^{18}$\,cm$^{-2}$ and
  $N$(H)$>3\cdot 10^{21}$\,cm$^{-2}$. Following \citet{crawford1985}, the
  opacity is more generally written as function of local density $n$
  and temperature $T$ of the \OI\ two level system:
  
  \begingroup
  \begin{eqnarray}
    \tau_0 = & \frac{\lambda^3\,A_{\rm{ul}}}{8\pi\Delta v_{\rm{FWHM}}} 
               \Bigl[(1+\frac{n_{\rm{cr}}}{n})\exp(228/T)-1\Bigr] \nonumber \\
    & \Bigl[\frac{\frac{3}{5}\exp(-228/T)}{1+\frac{3}{5}\exp(-228/T)+\frac{n_{\rm{cr}}}{n}}\Bigr] 
               \,\,N(\rm{OI})
    \label{eq:tauoi}
  \end{eqnarray}
  \endgroup

   \noindent with the Einstein A-coefficient
   $A_{\rm{ul}}=8.46\cdot 10^{-5}$\,s$^{-1}$, the critical density
   $n_{\rm{cr}}=4.7\cdot 10^{5}$\,\cmcub\ for collisions with H-atoms,
   $h\nu/k_B=228$\,K, and the ratio of statistical weights
   $g_u/g_l=3/5$. The resulting hydrogen column density for a line
   center opacity of the \OI\ 63$\mu$m line of 1 agrees within 10\%
   with the result from Eq.\,\ref{eq:oisimple} for $10^3$\cmcub\ $\le$
   n $\le$ $10^5$\cmcub\ and 20\,K\,$\le$\,400\,K. This is also
   consistent with the results of RADEX radiative transfer modeling
   \citep{vandertak2007}.

    We used the dust emission to estimate total hydrogen column
    densities at the positions observed in M\,33 by fitting a single
    modified black body (MBB) to the 70, 100, 160$\,\mu$m fluxes,
    while keeping $\beta=1.5$ constant, deriving the dust temperature
    and dust mass surface density. To find the best-fitting SED for
    each pixel on the map, a $\chi^2$ function was minimized using the
    Levenberg-Marquardt algorithm \citep{xilouris2012}.  For a
    constant gas-to-dust ratio of 150 \citep{kramer2010}, this implies
    typical hydrogen column densities of $2.4\cdot 10^{21}$\,cm$^{-2}$ per
    beam, implying moderate \OI\ opacities of 0.7.
    Hydrogen column densities peak at $\sim10^{22}$\,cm$^{-2}$
    ($A_V\sim10$\,mag) (Fig.\,\ref{fig:nhfromdust}), with
    corresponding \OI\ optical depths of $\sim3$.

   The standard PDR models of \citet{kaufman2006, pound2008}, used
   below, assume homogeneous slabs of an optical extinction of
   $A_V=10$\,mag. The models compute a simultaneous solution for the
   chemistry, the thermal balance, and also the radiative transfer. As
   already said above, optical depths effects of the emergent line
   emission are taken into account. From these models, it is known
   that the \OI(63$\mu$m) line emission becomes optically thick, with
   opacities of several, over the entire parameter space of $n$, $G_0$
   sampled by the models. These models do, however, not consider
   absorption of the emission of warm background gas by colder
   foreground gas, as would be expected for GMCs which are internally
   heated by star-formation.

   To estimate the possible effect of foreground absorption, we
   modeled the reduction in line center brightness temperatures
   assuming for each of the two source components beam filling factors
   of 1 and

     \begin{equation}
       T_{\rm R} = J_\nu(T_{\rm{ex}}) \,[1-\exp[-\tau_0]] 
     \end{equation}

     \noindent with $J_\nu(T_{\rm ex}) = 228\,(\exp(\frac{228}{T_{\rm
         ex}})-1)^{-1}$. The two source components are added, allowing
     for foreground absorption:

     \begin{equation}
        T_{\rm R} = T_{\rm{R,fg}} + \exp(-\tau_{\rm{0,fg}}) \,\, T_{\rm{R,bg}}.
     \end{equation}

     \noindent Using in addition Eq.\,\ref{eq:tauoi} for the opacity
     of the \OI\ line, we furthermore assumed for both components FWHM
     linewidths of 7\,\kms\ and a density of $n=10^4$\,\cmcub.  We
     assume that the background GMCs lie on average in the mid-plane
     of the galaxy and considered only half of the total column
     density in the following.  In the absence of high-resolution
     spectra, which would allow for fitting free parameters to the
     observed line profiles \citep[e.g.,][]{graf1993}, we assumed here
     ad-hoc that 80\% of this half total column density is in the
     foreground components at 15\,K, while 20\% is in a background
     component at $200$\,K. This results in a reduction of the
     emerging line intensity by a factor of 3.3 for the highest total
     column densities of $10^{22}$\,cm$^{-2}$ observed. The median
     corrections are $1.32\pm1.8$. The median corresponds to a total
     column density of $\sim1.17\cdot 10^{21}$\,cm$^{-2}$. A smaller
     fraction of foreground gas would lead to lower reduction factors
     and vice versa.
     %
     %
     We take these estimates as rough, first estimates and apply them to the
     data before comparing them with PDR models.

   Figure\,\ref{fig:ciioitir-observed} shows the observed intensity
   ratios and the ratios after correction of the \OI\ intensities.
   After the correction, the \CII/\OI(63$\,\mu$m) ratio of all
   positions ranges between 0.13 and 13.7 with a median of $3.0\pm2.1$,
   and the (\CII$+$\OI(63$\,\mu$m))/TIR ratio ranges between 0.3 and
   11.6\% with a median of $0.8\pm0.5$\%. The median ratios do not
   differ much between the different regions in M\,33
   (Table\,\ref{tab:cii-oi-tir}). 


\subsubsection{PDR models}

   To better understand the observed ratios of \CII/\OI\ and
   (\CII+\OI)/TIR, corrected for \OI\ self-absorption, we compare them
   with the predictions of PDR models
   \citep{kaufman2006,pound2008}. These assume FUV illuminated slabs
   of an optical exinction of $A_V=10$\,mag and solar metallicities, for
   a range of local densities and FUV fields.
 %
%
   The FUV field $G_0$ used in the PDR models is the local FUV field
   heating the slabs of dust and gas, leading to emission of the TIR,
   \CII, and \OI. The ratio of the local $G_0$ from the models and the
   \gobs\ from the observed TIR emission gives the beam filling factor
   of the emitting clouds. Beam filling factors in the observed ratios
   cancel out under the reasonable first order assumption that the
   three tracers (\CII, \OI, TIR) have the same filling factor. We
   ignore here that the excitation conditions for the \OI\ line
   indicate a somewhat smaller beam filling factor than for the other
   two tracers.

   For many of the observed ratios, the PDR models do not provide a
   unique $G_0,n$ solution. For these positions, two solutions exist,
   one at high density and low FUV field, and another one at moderate
   values of $G_0$ and $n$. An example is shown in
   Figure\,\ref{fig:kaufman-all}, which exhibits contours of the
   average ratios of all observed positions
   (Table\,\ref{tab:cii-oi-tir}) as a function of density and FUV
   field of the Kaufman PDR models. There exist two best-fitting
   solutions, which are both consistent with the ratios of \CII, \OI,
   and TIR.

%
%
%

Solutions with moderate $n$, $G_0$ exist for the bulk of the ratios in
M\,33, after correcting for \OI\ self-absorption
(Fig.\,\ref{fig:ciioi}\,Upper panel). The ratios span the range of
$\log(G_0)<3$ and $1\la\log(n/$\cmcub$)\la4.25$. The average ratios of
all four regions can be explained by FUV-fields of $n\sim2\cdot
10^2$\,\cmcub\ and $G_0\sim60$ (Fig.\,\ref{fig:kaufman-all},
Table\,\ref{tab:cii-oi-tir}). The observed FUV fields \gobs\ range
between 2 and 200, which indicates beam filling factors of about 1.
The average ratios of the four regions do not differ much and lead to
similar best fitting $n$, $G_0$ solutions. Even after correcting the
\OI\ intensities, some of the ratios still lie outside of the
parameter space spanned by the moderate solutions: for example the
ratios with (\CII+\OI)/TIR $\sim0.8\%$ and \CII/\OI$>4$. Corrections
of the \CII\ intensities for contributions from the diffuse, ionized
gas may, however, move all points to lower \CII/\OI\ ratios and lower
(\CII+\OI)/TIR rations, and into this parameter space.

The low-FUV solutions of the PDR models do, on the other hand, cover
the entire range of ratios (Fig.\,\ref{fig:ciioi}\,Lower panel). The
ratios lie in the regime $0\la\log\,G_0\la0.75$ and
$3.5\la\log(n/$\cmcub$)\la4.5$.  The average of all ratios is best
fit by $n\sim10^4$\,\cmcub, $G_0\sim1.5$
(Fig.\ref{fig:kaufman-all}, Table\,\ref{tab:cii-oi-tir}). Given the
range of observed \gobs, this would indicate beam filling factors
$\Phi_{\rm{G0}}$ of between $\sim1$ for the regions with the lowest
TIR and $\sim100$ for the active arm regions with the highest observed
TIR. While a fraction of the PDRs along the lines-of-sight may be
represented by the low-FUV solution, this cannot be the sole
solution. Such a high number of PDRs along the lines-of-sight seems
not consistent with the observed peak optical extinctions of
$A_V\sim10$\,mag, which resemble those of a single PDR model slab.
However, only the surface columns of these models of about 1\,mag, or
of a hydrogen column density of $\sim10^{21}$\,cm$^{-2}$, emit
\CII\ (cf. Fig.\,2 in \citet{kaufman1999}) and much of the remaining,
deeper layers may not exist. High filling factors also seem not
consistent with the relatively short lines-of-sight in M\,33 with its
moderate inclination of only $56\degr$. Comparing the observed
\CII\ intensities with those predicted by the PDR models also gives
high beam filling factors for the low-FUV solutions.
Table\,\ref{tab:cii-oi-tir} lists the ratios for a position with
particularly strong \CII\ intensity. The best fitting low-FUV solution
for this position is $n=5\cdot 10^3$\,\cmcub, $G_0=3$
(Fig.\ref{fig:kaufman-all}). For this solution, the ratio of observed
to modeled \CII\ intensity is $\Phi_{\rm{CII}}\sim10$, which again
seems difficult to reconcile with the models.  For the moderate
solution at the \CII\ peak position, on the other hand, $\Phi_{\rm
  CII}$ is $\sim1$.


 The degeneracy of the PDR model solutions has been discussed, for
 example by \citet[][]{hughes2015, parkin2013, kramer2005,
   higdon2003}, and \citet{malhotra2001} for a variety of normal
 galaxies, who all prefer the moderate solution, often argueing that
 the low-FUV solution would require far too many PDR slabs along the
 lines-of-sight to be consistent with the observed intensities.

 However, it is also clear that GMCs and their OB associations
 illuminating gas and dust show structure on a wide range of scales,
 and that the PDR model slab of constant volume and column density,
 illuminated by a constant FUV field is only a first order
 approximation. The inner parts of the GMCs and their OB associations
 and \HII\ regions are illuminated by high FUV-fields, while the
 less dense, colder, outer, and more remote, diffuse parts of the 
 GMCs are illuminated by much lower FUV-fields. We therefore
 cannot exclude that both solutions of the PDR models are at play
 along the lines-of-sight in M\,33.

 Detailed modeling of the spectral energy distributions (SEDs) may
 shed light on the relative fractions of gas and dust which exhibit
 the two solutions of the Kaufman PDR models. \citet{aniano2012}
 define PDRs as those reagions which are heated by starlight
 intensities which are a factor 100 or more higher than the ISRF in
 the solar neighborhood \citep{mathis1983}\footnote{The
   \citet{mathis1983} estimate for the ISRF has $G_0=1.14$
   \citep[e.g.,][]{draine2011}.}, $U>100$, tracing massive star forming
 regions heated by OB stars. They use \citet{draine-li2007} dust
 models, which assume a distribution of radiation fields between a
 minimum and a maximum value, fitting them to the observed SEDs
 between 3.6\,$\mu$m and 500\,$\mu$m at each position of the galaxies
 NGC\,628 and NGC\,6946, creating maps of the PDR fraction. They find
 that only a fraction of 12\% to 14\% of the TIR emission stems from
 such PDRs. More than 85\% of the TIR emission in these two galaxies
 stems from the diffuse ISM heated by low starlight intensities. The
 latter regions may resemble the regions in M\,33 which are
 characterized by the low-FUV solution of the Kaufman PDR models.

\begin{figure}[]
  \includegraphics[angle=0,width=0.48\textwidth]{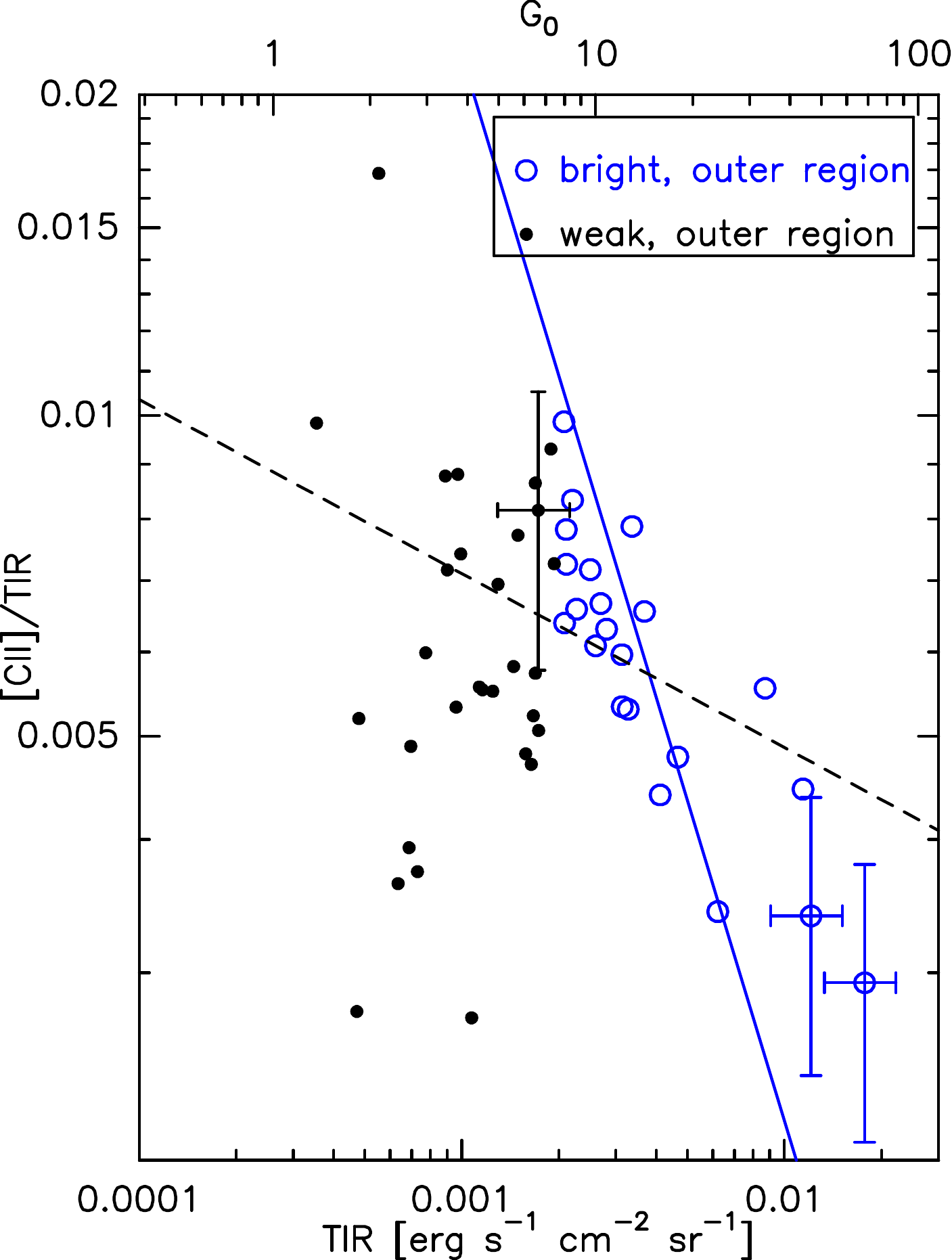}
  \caption{\CII/TIR intensity ratio versus TIR for the northern
    \HII\ region BCLMP\,691 in M\,33. The blue drawn line shows the
    result of an unweighted linear least-squares fit to the TIR-bright
    data of BCLMP\,691 varying only the y-intercept, while keeping the
    slope fixed at $-1$. The black dashed line shows the result of an
    unweighted linear least-squares fit to all M\,33 data
    (cf. Fig.\,\ref{fig:ciitir-tir-r}). A few typical 1$\sigma$
    errorbars are shown. }
  \label{fig:cii-tir-bclmp691}
\end{figure}

\subsection{Metallicities}


To first order, one may expect that low metallicities lead to a rise
of \CII/TIR ratios, as a lowered dust abundance leads to deeper
penetration lengths of UV photons, increasing the \CII\ surface layers
\citep{bolatto1999}. This may increase \CII\ intensities, while
the thermal dust flux stays about constant \citep{israel1996}. 
Indeed, low metallicities have been proposed to explain the observed
high \CII/TIR ratios found for example in dwarf galaxies \citep{cormier2015}
and in the outskirts of spirals \citep[e.g.,][]{kapala2015,kapala2017}.

M\,33 exhibits about half solar metallicities together with shallow
gradients of dropping metallicities with increasing galacto-centric radius
\citep{magrini2010}. \citet{toribio-san-cipriano2016} measured radial
abundance gradients of O/H and C/H from observations of \HII\ regions
in M\,33.
From optical recombination lines they find:

\begin{eqnarray}
12+\log({\rm O/H}) & = & 8.76 -(0.33\pm0.13) \times R/R_{25} \\
12+\log({\rm C/H}) & = & 8.64 -(0.61\pm0.11) \times R/R_{25}.
\label{eq:metallicitygradient}
\end{eqnarray}

The value of the 25th mag B-band isophotal radius ($R_{25}$) is
$28$\,arcmin (6.84\,kpc).
This is the first work to discuss the C/H abundance gradient of M\,33
based on more than just two \HII\ regions \cite[cf. the discussion
  in][]{toribio-san-cipriano2016}. The observed C/H gradient is about
twice as steep as the O/H gradient, similar to what is found in other
small galaxies with subsolar metallicities.
%
%
The observed metallicity gradient may contribute to
  the observed indications of a rise of the \CII/TIR ratio in the
  outer regions of M\,33 (Fig.\,\ref{fig:ciitir-tir-r}).
%
%
However, the scatter of the observed \CII/TIR ratio at
  a given radial distance in M\,33 (Fig.\,\ref{fig:ciitir-tir-r}) and
  hence about constant metallicity, indicates that other mechanisms
  can become important. 
%
%

Here, we take as one example the \HII\ region \object{BCLMP\,691},
which is located at a galacto-centric distance of 3.3\,kpc in the far
north of M\,33 \citep{braine2012} at about constant metallicity
(Eq.\,\ref{eq:metallicitygradient}).  For a small region of
$\sim0.9$\,arcmin$^2$ above the TIR threshold, BCLMP\,691 exhibits a
marked drop of the observed \CII/TIR ratio from the outskirts of the
\HII\ region to its center and the peak of TIR emission by more than a
factor of 3, from $\sim1$\% to $0.3$\% (Figs.\,\ref{fig:strips-north},
\ref{fig:cii-tir-bclmp691}).

For optically thin emission, the TIR equals

\begin{equation}
  {\rm TIR} = \int B_\nu(T_d) \tau_d d\nu = \Sigma_d \kappa_0 \int
  B_\nu(T_d) (\nu/\nu_0)^\beta d\nu
\end{equation}

with the total dust mass surface density $\Sigma_d$, the line-of-sight
(l.o.s.) weighted averaged dust temperature $T_d$, the Planck function
$B_\nu$, the opacity $\tau_d$, the average grain cross-section per
gram $\kappa_0$ at frequency $\nu_0$, and the l.o.s. averaged dust
emissivity index $\beta$, assuming optically thin emission.

Figure\,\ref{fig:cii-tir-bclmp691} shows the observed $\log$\CII/TIR
vs. $\log$TIR in BCLMP\,691. A linear least squares-fit to the
positions above the TIR threshold (Fig.\,\ref{fig:strips-north}),
keeping the slope fixed to $-1$ results in a correlation coefficient
of $r=-0.84$, indicating a good correlation. This slope is consistent
with a constant slab of \CII\ emission of
$2.2\cdot 10^{-5}$\,\ergintensity, the dashed contour in the corresponding
\CII\ map (Fig.\,\ref{fig:strips-north}).
The \CII\ map indeed exhibits a roughly constant emission over the
TIR-bright region, indicating a constant column density and excitation
temperature. MBB fits to the SEDs within the small area around the
peak of TIR emission in BCLMP\,691, where TIR intensities are above
the threshold, show that dust temperatures stay fairly constant
($23.2\pm1.9$\,K) with a median value of the estimated errors of
$T_{\rm{dust}}$ of 1.3\,K. On the other hand, the dust mass surface
densities vary by a factor of $\sim4$ between 200 and 800\,\msun/beam
(Fig.\,\ref{fig:bclmp691-md-td}), a variation which is a factor 3.5
larger than the median of the estimated errors, which is
170\,\msun/beam. The observed steep drop of \CII/TIR with TIR in a
region of about constant metallicity is hence naturally explained.

\section{Summary and conclusions}

The emission lines of \CII\ and \OI(63$\mu$m) were mapped along the
major axis of the Local Group galaxy M\,33. These maps have a width of
$\sim370$\,pc and cover a region of 38\,arcmin$^2$ at 50\,pc
resolution, allowing to resolve arm and interarm regions.  The
southern most region lies at 2\,kpc galacto-centric distance and,
located at the other side of the galaxy, the northern most region lies
at 5.7\,kpc distance. These maps much improve on the 1-dimensional
\CII\ cut at 280\,pc resolution, which had been observed along the
major axis of M\,33 using ISO/LWS \citep[][]{kramer2013}.

We combined full-galaxy maps at 24$\mu$m, 70$\mu$m, 100$\mu$m, and
160$\mu$m to construct a map of the total infrared continuum emission
(TIR), integrated between 1$\mu$m and 1000$\mu$m wavelength using the
kernels provided by \citet{aniano2011} and the weighting factors
derived by \citet{boquien2011}. The observed range of TIR intensities
translates to a range of FUV fluxes of \gobs$\sim 2$ to 200 in units
of the average Galactic radiation field.

We find that the TIR and \CII\ intensities are tightly correlated over
two orders of magnitude.  The average \CII/TIR ratio of
$0.64\pm0.23$\% is not significantly higher than the average \CII/TIR
ratio of $0.48\pm0.21$\% found in a sample of 54 nearby galaxies by
\citet{smith2017}. \CII/TIR ratios observed in the two northern most
regions at 4.5 and 5.5 galacto-centric distances show increasing
average ratios of 0.8 and 1.3, respectively. The large-scale variation
of the \CII/TIR ratios is consistent with the ISO/LWS observations.
The resolution and extent of the PACS map allows to distinguish
between the diffuse inter-arm regions of the inner and outer galaxy,
and the arm regions.
%
%
The \CII/TIR ratio averaged over bins of 0.5\,dex decreases with
increasing TIR from $1.1\pm0.4$\% for regions with weak TIR emission
to $0.5\pm0.1$\% for the arm-regions with highest TIR, while the
scatter of binned averages of the \CII/TIR ratios decreases as well.

The drop of \CII/TIR ratios toward sites of massive star forming
regions, where TIR peaks, is most clearly visible in the \CII/TIR map
of one of the northern \HII\ regions, BCLMP\,691, at 3.3\,kpc
galacto-centric distance, for positions where TIR is bright. In this
case, the drop of \CII/TIR ratios is consistent with a \CII\ surface
layer of constant intensity, which is independent of TIR. The rise of
TIR is caused by a rise of dust mass surface densities, for about
constant dust temperatures and emissivities, as is confirmed by the
results of modified black body (MBB) fits. For this \HII\ region, the
observed steep drop of \CII/TIR with TIR in a region of about constant
metallicity is hence naturally explained. However, this does not rule
out that on larger scales the drop of metallicities with
galacto-centric distance observed in M\,33 \citep{magrini2010} is
decisive to explain the observed drop of \CII/TIR with TIR on these
scales.



In M\,33, \CII\ intensities are stronger than those of the
\OI(63$\mu$m) line at almost all positions. The \CII/\OI\ ratio varies
between $\sim0.2$ and 20, with a mean of $4.5\pm2.6$. At first glance,
the \OI\ and \CII\ maps of the inner galaxy resemble each
other. However, closer inspection shows that \OI\ and \CII\ are only
correlated in the TIR-brightest regions while there is only little
correlation of both tracers in the TIR weak regions. The observed
(\CII$+$\OI)/TIR ratio varies between $\sim0.3\%$ and 3\%, with an
average of $0.75\pm0.3$\%.

Data of atomic hydrogen, taken with the VLA at the same resolution as
the \CII\ data, were used to estimate the contribution of the cold
neutral medium (CNM) to the observed \CII\ emission, following the
approach of K2013. As anticipated from this work, the averaged
contribution is only $\sim10$\%. We did not correct the \CII\ emission
for this minor contribution. Modified black bodies were fit to the
continuum emission at 70, 100, 160\,$\mu$m to derive dust temperatures
and surface densities, which were used to estimate total gas and O
column densities, and the optical depths of the \OI(63$\,\mu$m)
line. A simple model of cold foreground and warm background gas then
allowed us to estimate that for the highest \HI\ colunn densities of
$10^{22}$\,cm$^{-2}$, \OI(63$\,\mu$m) line intensities are reduced by
a factor of 3.3, while the median correction factor is
$1.3\pm1.8$. The \OI\ data were corrected for this effect. An
additional correction of the \CII\ data, for diffuse, ionized gas, was
not attempted here.

The observed \CII/\OI\ and (\CII+\OI)/TIR ratios were corrected for
\OI\ foreground absorption, and then compared with standard PDR models
\citep{kaufman2006}. Averages of all observed ratios are similar to
the averages of the four individual regions. They all show two
solutions of the PDR models, a moderate solution with $n\sim2\cdot
10^2$\,\cmcub, $G_0\sim60$, and a low-FUV solution with
$n\sim10^4$\,\cmcub, $G_0\sim1.5$. The bulk of the observed positions
can be modeled by a moderate solution. This solution implies low beam
filling factors of $\sim1$. The low-FUV solution, on the other hand,
cannot be the sole solution for all gas along the lines of sight, as
it would imply very high beam filling factors $\gg1$, which are
inconsistent with the observed FUV fields, the \CII\ intensities, and
the total column densities.


  
\begin{acknowledgements}
  We thank the anonymous referee for insightful comments which helped
  to improve the paper, Mark Wolfire and Alessandra Contursi for
  helpful discussion, and the NHSC team at IPAC for their support in the
  data reduction process.
  %
  %
  M.R. and S.V. acknowledge support by the research projects
  AYA2014-53506-P and AYA2017-84897-P from the Spanish Ministerio de
  Econom\'{i}a y Competitividad, from the European Regional
  Development Funds (FEDER) and the Junta de Andaluc\'{i}a (Spain)
  grants FQM108. This study has been partially financed by the
  Consejer\'{i}a de Conocimiento, Investigaci\'{o}n y Universidad,
  Junta de Andaluc\'{i}a and European Regional Development Fund
  (ERDF), ref. SOMM17/6105/UGR. FST thanks the Spanish Ministry of
  Economy and Competitiveness (MINECO) for support under grant number
  AYA2016-76219-P.
\end{acknowledgements}

\bibliography{m33} 


\begin{appendix}


\section{Observing parameters of PACS spectroscopy observations}

\begin{table*}
  \caption{Observing parameters of all 22 fields mapped in M\,33 with
Herschel/PACS in line spectroscopy mode.  }
\label{tab:obssum}
\centering
\begin{tabular}{l c c c c c}
\hline\hline
Target-Name & RA & Dec. & Integr. Time & ObsID & Obs. Day \\
            & (eq2000)  &   & (sec) & & \\
\hline
M33-1350N  &  1h34m31.15s  &  +31d00m23.90s  &  6777  &  1342213044  &  616 \\
M33-1150N  &  1h34m25.19s  &  +30d57m19.20s  &  6779  &  1342213043  &  616 \\
M33-800N   &  1h34m14.80s  &  +30d51m55.80s  &  6781  &  1342213045  &  616 \\
M33-600N   &  1h34m08.80s  &  +30d48m51.00s  &  6779  &  1342213046  &  616 \\
M33-500N   &  1h34m06.12s  &  +30d47m19.26s  &  3876  &  1342189072  &  238 \\
M33-400N   &  1h34m02.90s  &  +30d45m46.20s  &  6777  &  1342212583  &  609 \\
M33-350N   &  1h34m01.42s  &  +30d45m00.10s  &  6777  &  1342212582  &  609 \\
M33-250N   &  1h33m58.45s  &  +30d43m27.70s  &  6777  &  1342212579  &  609 \\
M33-200N   &  1h33m57.00s  &  +30d42m41.50s  &  6777  &  1342212540  &  608 \\
M33-150N   &  1h33m55.48s  &  +30d41m55.30s  &  6775  &  1342212538  &  608 \\
M33-100N   &  1h33m54.00s  &  +30d41m09.10s  &  6775  &  1342212536  &  608 \\
M33-50N    &  1h33m52.51s  &  +30d40m22.90s  &  6775  &  1342212534  &  608 \\
M33        &  1h33m51.02s  &  +30d39m36.70s  &  6775  &  1342213047  &  616 \\
M33-50S    &  1h33m49.54s  &  +30d38m50.60s  &  6775  &  1342212535  &  608 \\
M33-100S   &  1h33m48.10s  &  +30d38m04.30s  &  6775  &  1342212537  &  608 \\
M33-150S   &  1h33m46.58s  &  +30d37m18.20s  &  6773  &  1342212539  &  608 \\
M33-200S   &  1h33m45.10s  &  +30d36m31.90s  &  6773  &  1342212578  &  609 \\
M33-250S   &  1h33m43.61s  &  +30d35m45.80s  &  6773  &  1342212580  &  609 \\
M33-300S   &  1h33m42.10s  &  +30d34m59.50s  &  6773  &  1342212581  &  609 \\
M33-400S   &  1h33m39.20s  &  +30d33m27.20s  &  6773  &  1342212584  &  609 \\
M33-450S   &  1h33m37.68s  &  +30d32m41.00s  &  6773  &  1342212585  &  609 \\
M33-500S   &  1h33m36.20s  &  +30d31m54.80s  &  6773  &  1342212586  &  609 \\

\hline
\end{tabular}
\end{table*}

\section{Figures}

\begin{figure*}[h!]
  \centering
  \includegraphics[width=0.49\textwidth]{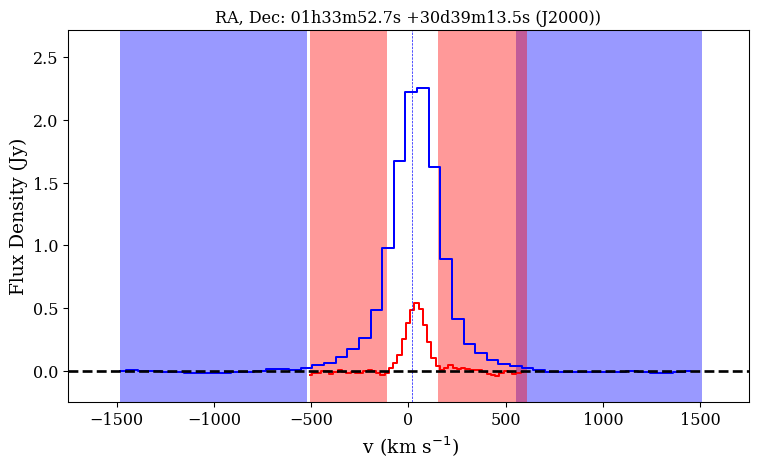}
  \includegraphics[width=0.49\textwidth]{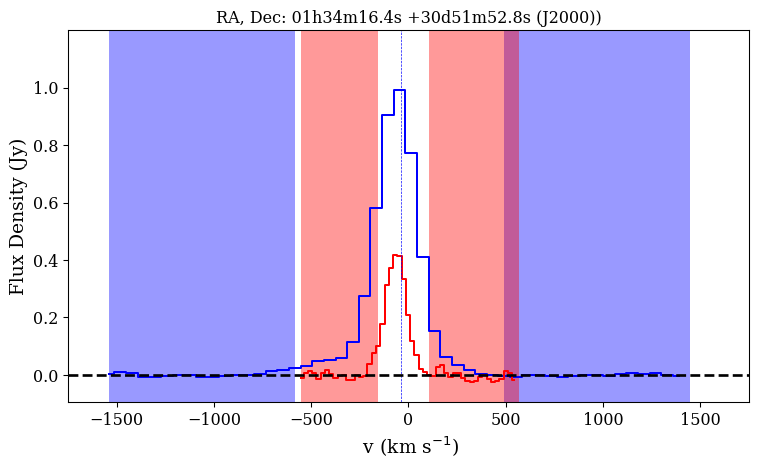} \\
  \includegraphics[width=0.49\textwidth]{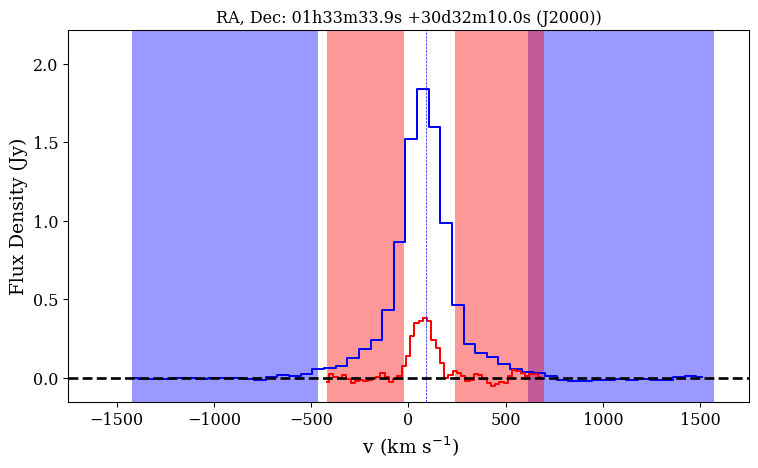}
  \includegraphics[width=0.49\textwidth]{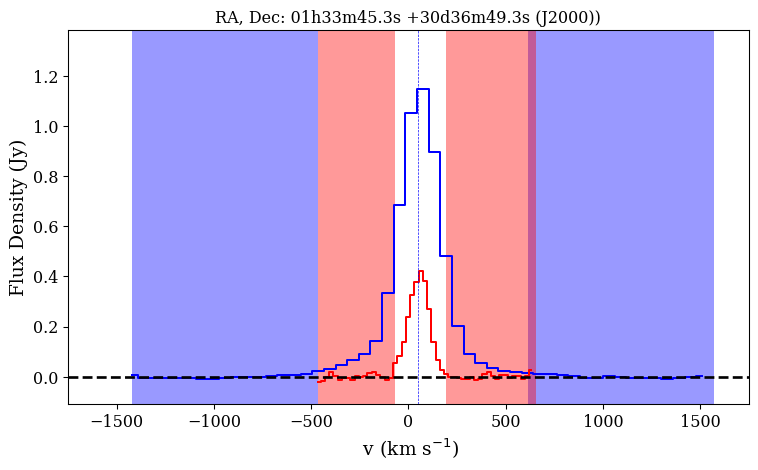} \\
  \includegraphics[width=0.49\textwidth]{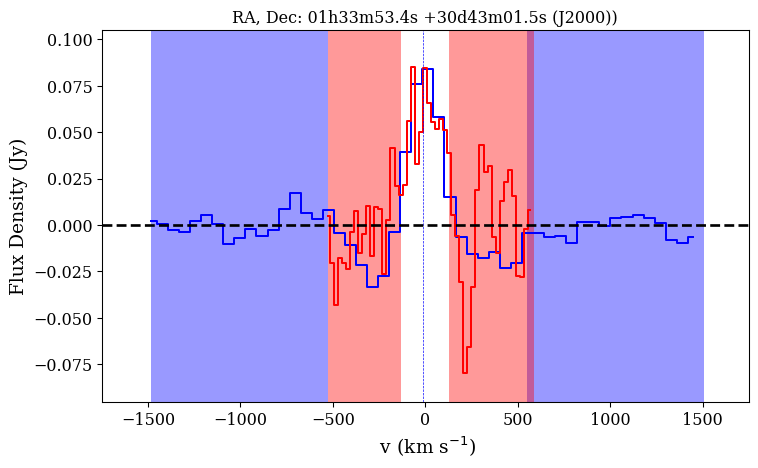}
  \includegraphics[width=0.49\textwidth]{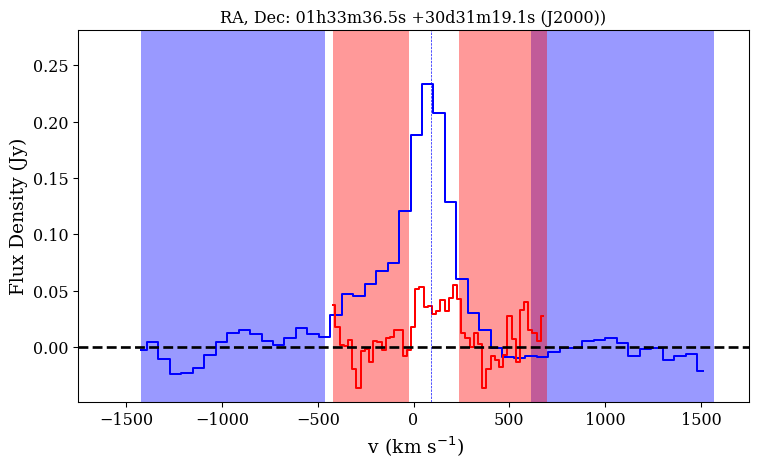} \\
%
%
  \caption{ PACS spectra of \CII\ (blue) and \OI(63$\mu$m) (red) at
    selected positions in M~33. All spectra are shown at their native
    angular resolution, before convolution to $12''$ resolution. The
    velocities have been corrected for the systemic velocity of M\,33
    of $-180$\,\kms. Dashed, vertical lines mark the residual velocity
    of the \HI\ line \citep{warner1973}. Spectra are shown after
    subtraction of polynomial baselines of up to 3rd order. The two
    red (blue) fields mark the velocity range, which was used to
    calculate the baseline rms for the \OI\ (\CII) line. Baselines at
    more extreme velocities are not shown. \CII\ integrated
    intensities were calculated over the inner range between the blue
    fields ($\pm538$\,\kms), and the integrated intensities for the
    \OI\ line were calculated over the spectral range (in white)
    between the red fields.  The double-peak structure of the weak
    \OI\ line seen in the right spectrum of the lower row is
    attributed to baseline noise. }
\label{fig:spectra}
\end{figure*}

\begin{figure}[h!]
\includegraphics[angle=0,width=0.48\textwidth]{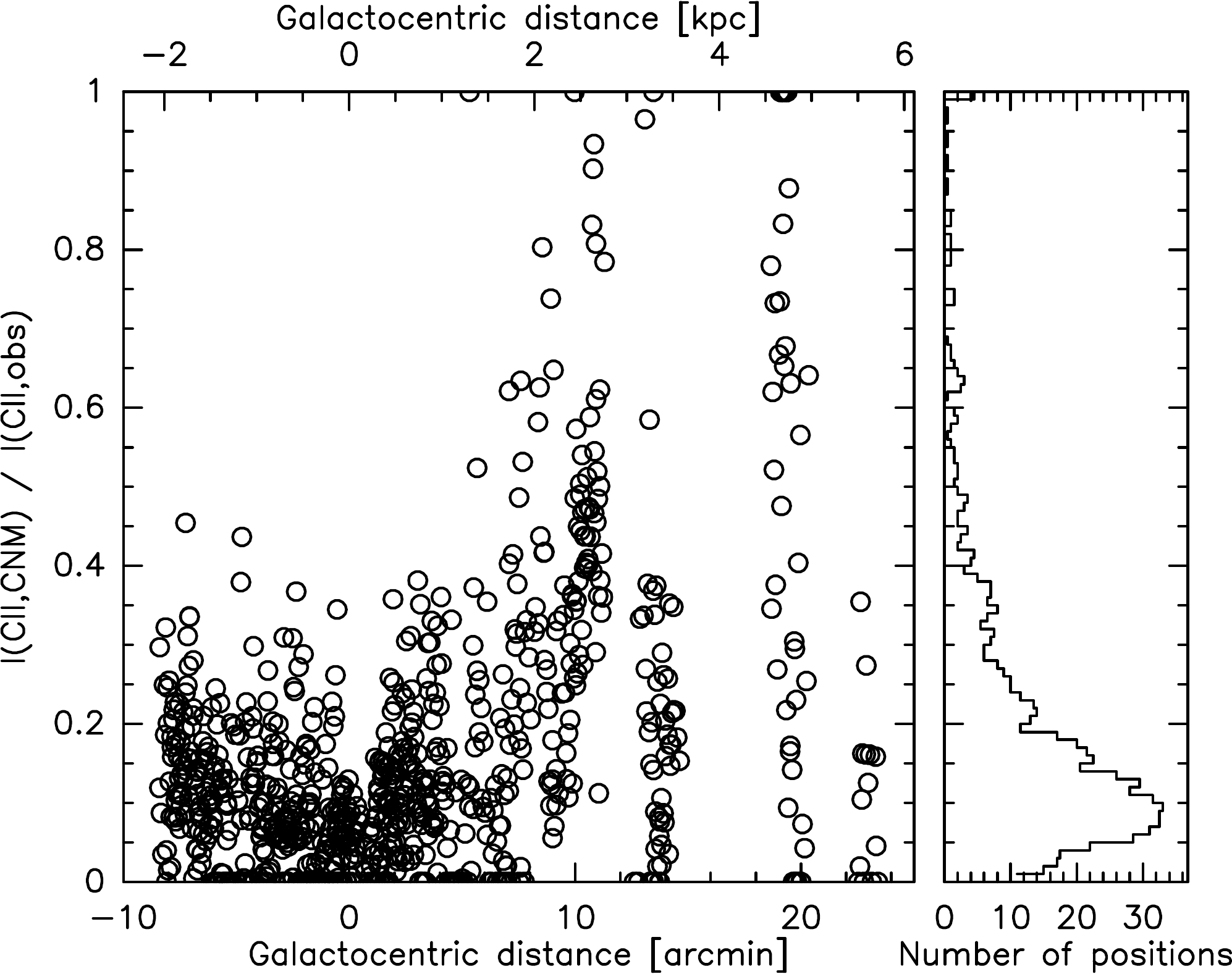}
\caption{Fraction of \CII\ emission from the cold neutral medium
  (CNM), estimated from \HI\ emission.}
 \label{fig:iciicnmfrac} 
\end{figure}

\begin{figure}[h!]
\includegraphics[angle=0,width=0.48\textwidth]{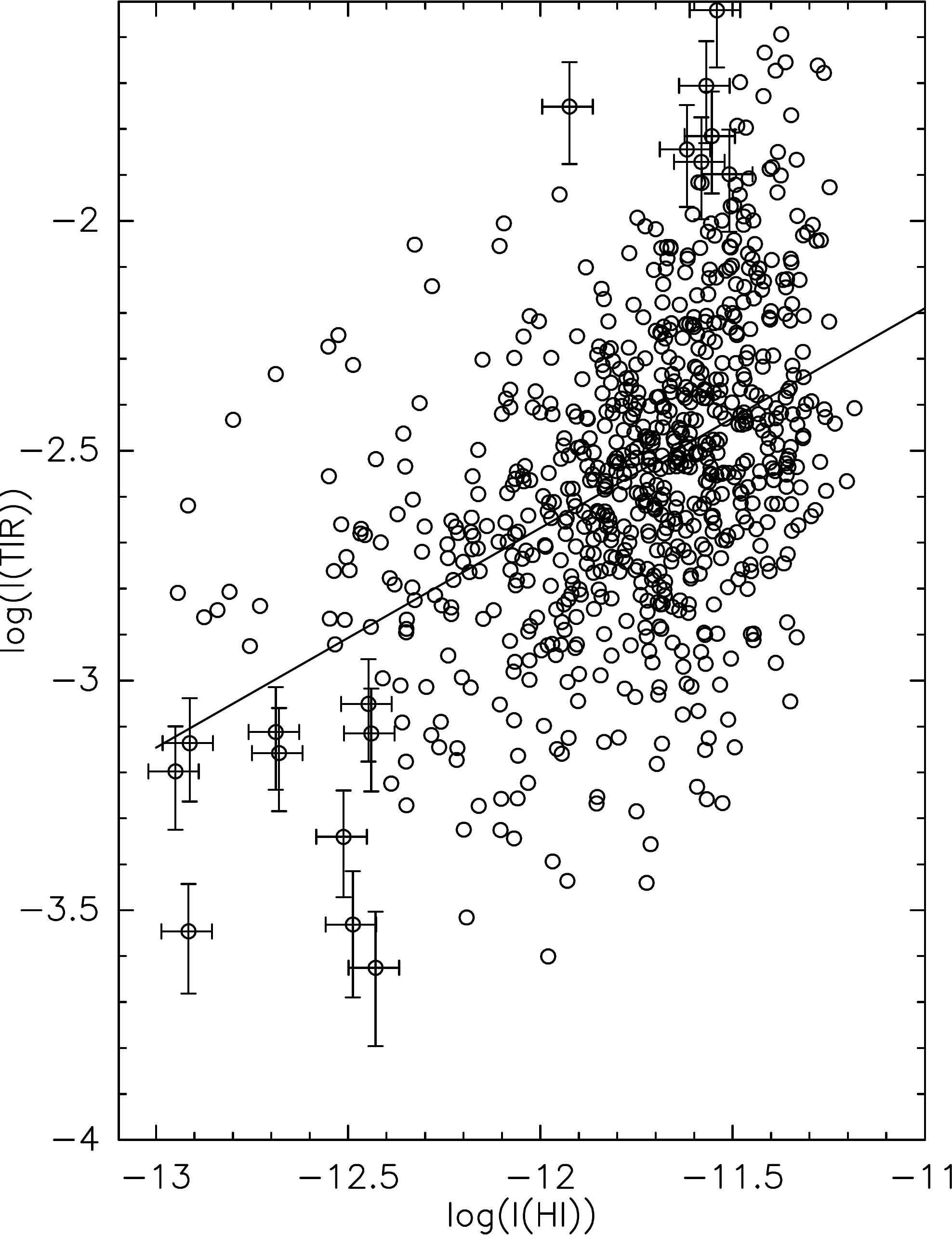}
\caption{TIR versus \HI\ intensities in M\,33. Intensities are given
  in \ergintensity. Typical $1\sigma$ errorbars are shown. For \HI\ we
  assume an calibration error of 15\% (cf. K2013). The drawn line
  shows the result of an unweighted linear least-squares fit to
  $\log\,{\rm TIR}=b+m\times\log\,{\rm HI}$, which gives a slope
  $m=0.48\pm0.11$, y-intercept $b=3.07\pm1.32$, and correlation
  coefficient $r=0.45$.}
 \label{fig:tir-hi} 
\end{figure}

\begin{figure}[h!]
 \includegraphics[angle=0,width=0.48\textwidth]{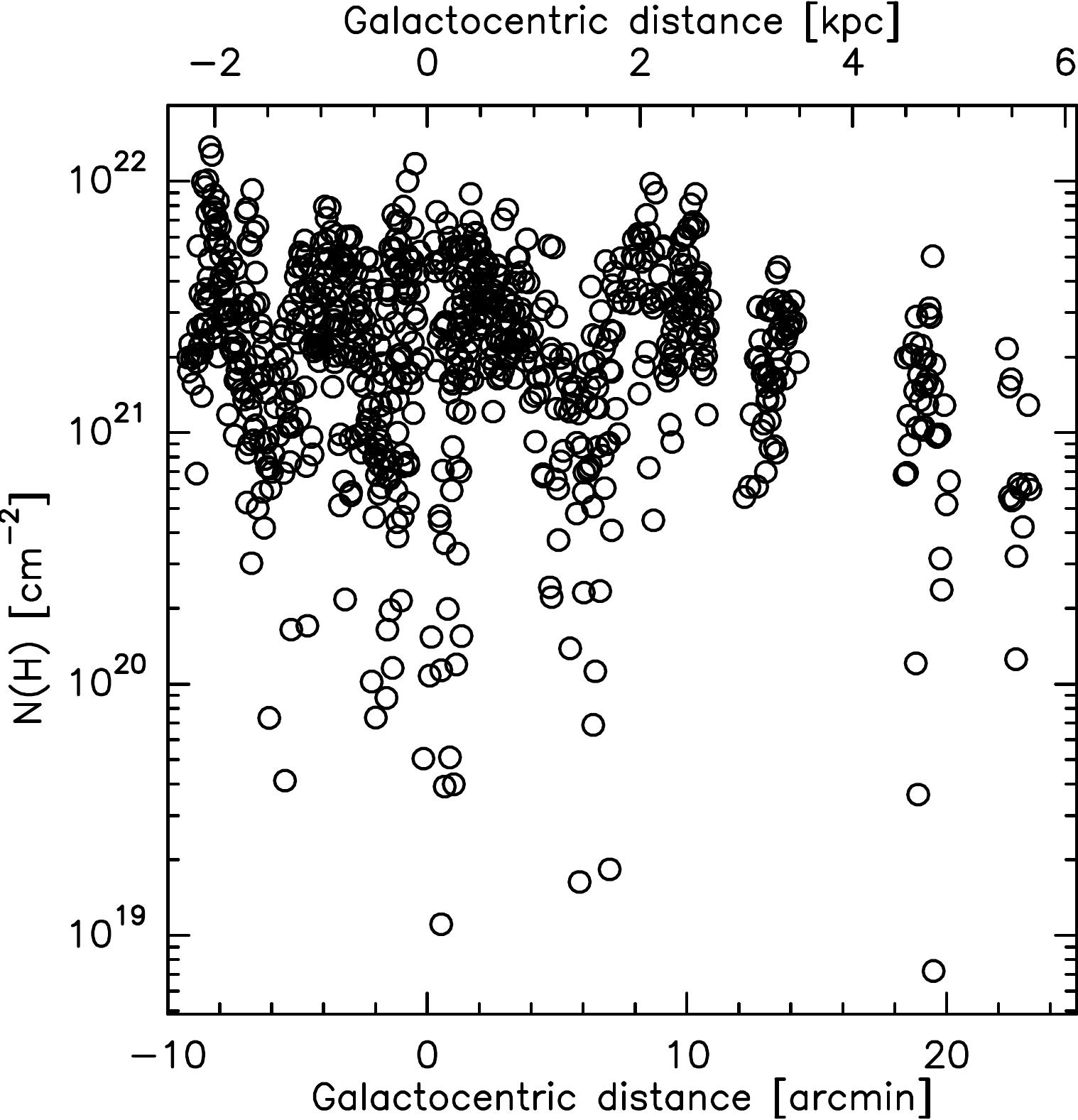}
 \caption{Total hydrogen column densities derived from fits of
   modified black bodies to the $70\,\mu$m, $100\,\mu$m, and
   $160\,\mu$m data at each position, assuming a constant gas-to-dust
   ratio (see Sec.\,\ref{sec:oicorr}).}
 \label{fig:nhfromdust}
\end{figure}

\begin{figure}[h!]
 \includegraphics[angle=0,width=0.48\textwidth]{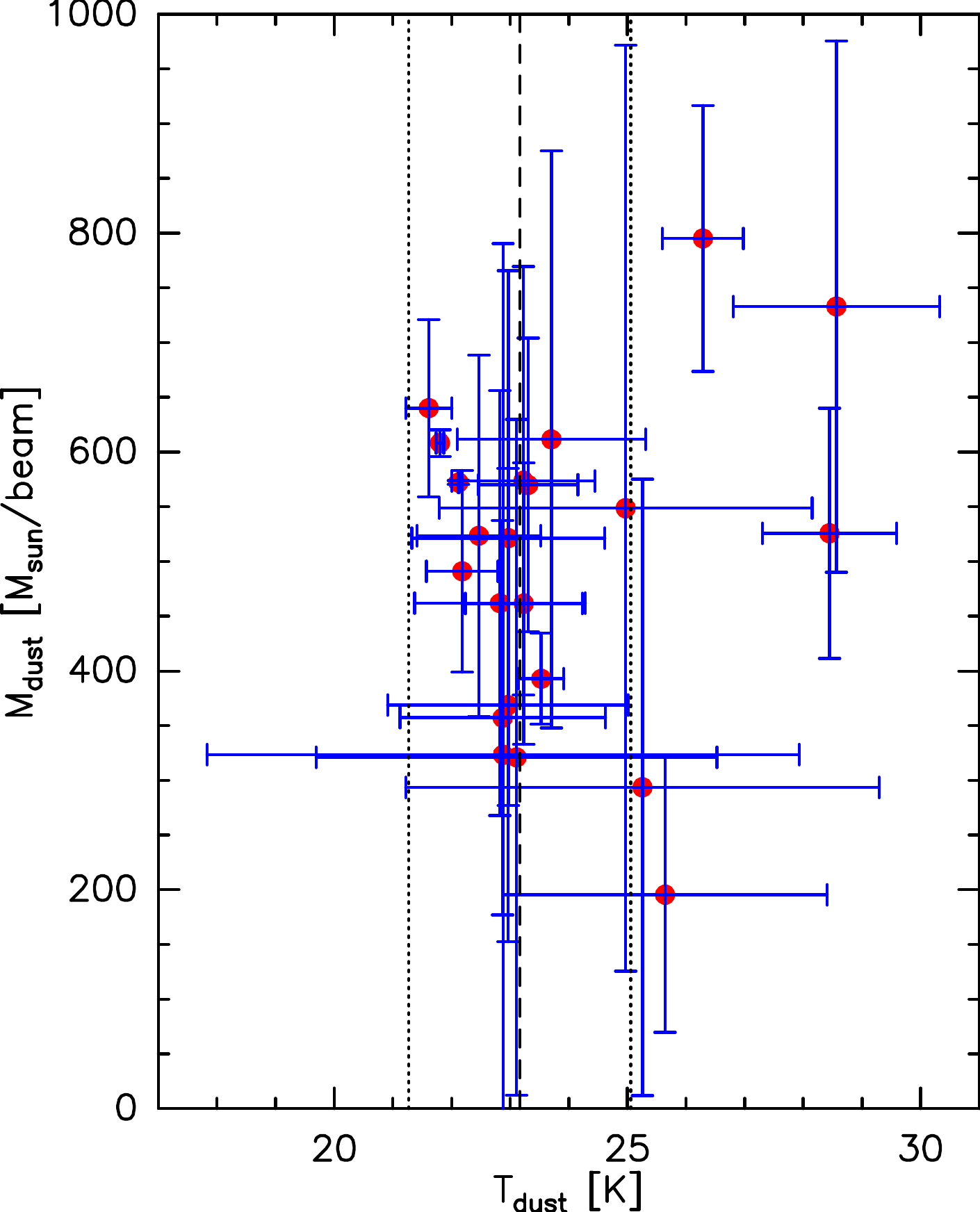}
 \caption{Dust mass surface density vs. dust temperature, derived from
   fits of modified black bodies, for the TIR bright positions in the
   \HII\ region BCLMP\,691. The thick dashed line shows the median
   dust temperature and thin dashed lines correspond to the rms
   scatter.}
 \label{fig:bclmp691-md-td}
\end{figure}

\begin{figure*} 
  \includegraphics[angle=0,width=0.55\textwidth]{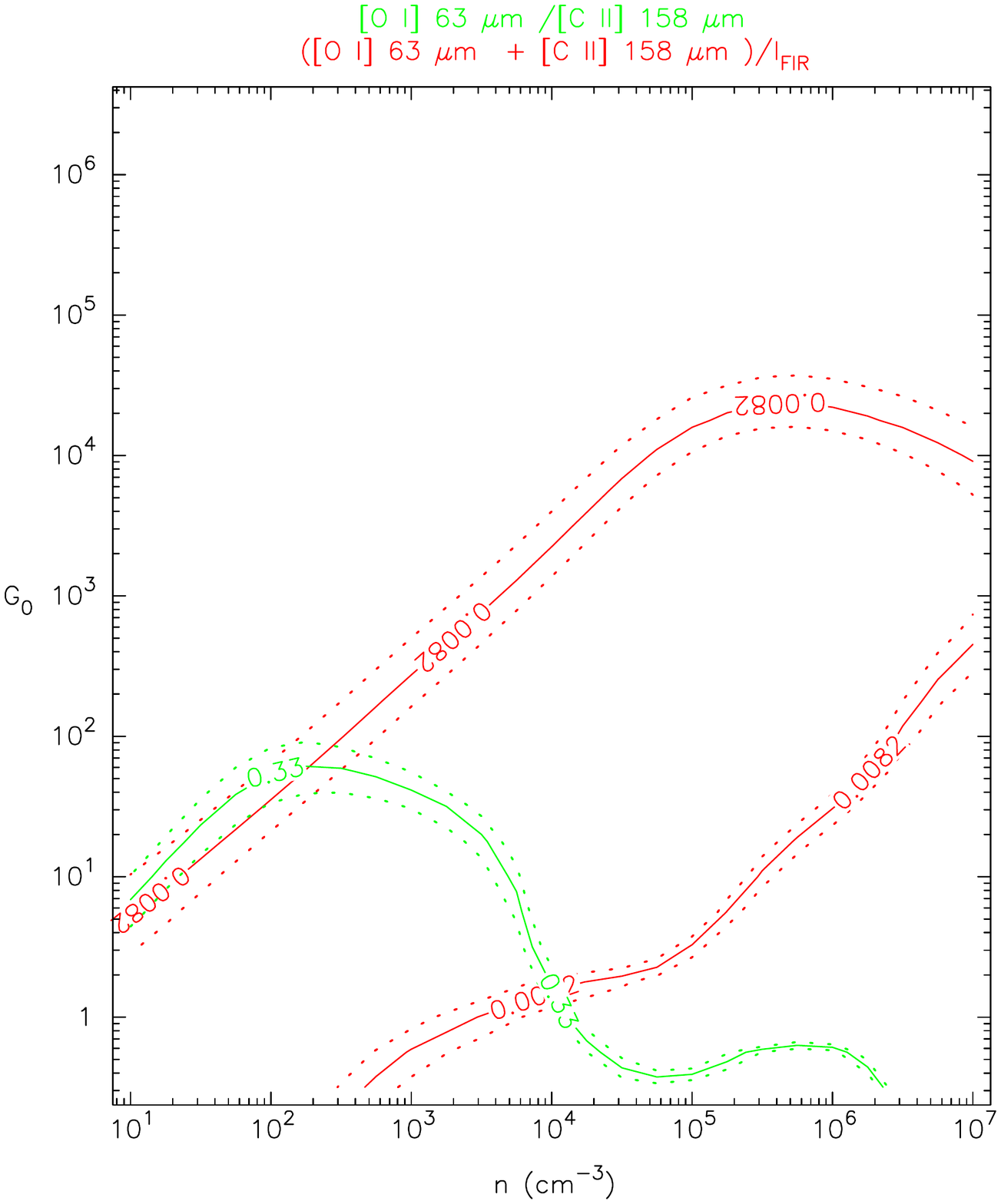} \hspace{-1.5cm}
  \includegraphics[angle=0,width=0.55\textwidth]{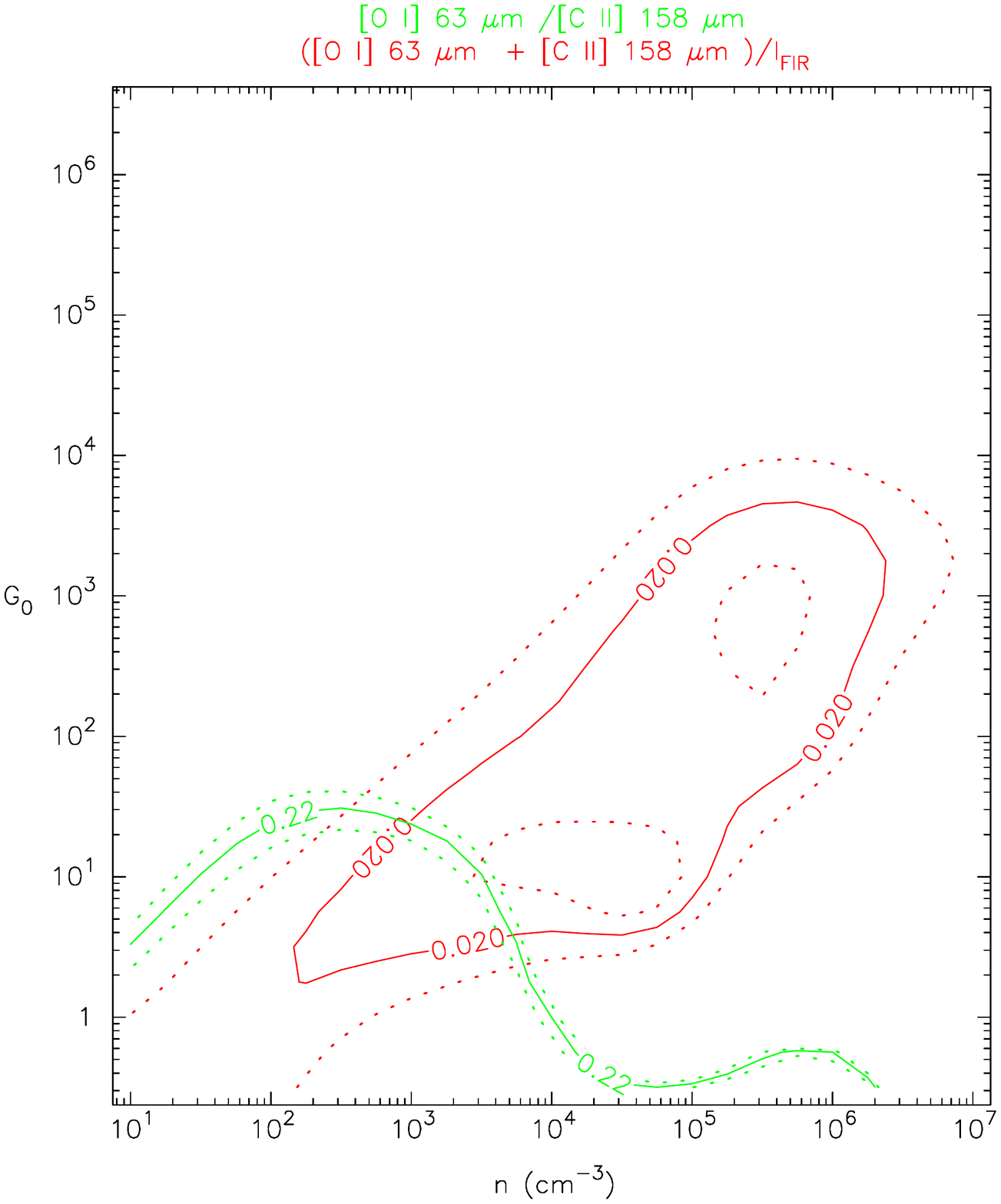} 
 \caption{Contours of the ratios of \OI/\CII, and (\CII+\OI)/TIR, with
   \OI\ corrected (Table\,\ref{tab:cii-oi-tir}), as a function of
   cloud density $n$ and incident FUV flux $G_0$ using the PDR model
   of \citet{kaufman2006}. {\bf Left:} Median ratios of all
   positions. {\bf Right:} Ratios at a \CII\ peak position. }
 \label{fig:kaufman-all} 
\end{figure*}

\end{appendix}


\end{document}